\documentclass{aa}
\usepackage{amssymb}
\usepackage{amsmath}
\usepackage{txfonts}
\usepackage{graphicx}
\usepackage{natbib}
\usepackage{cancel}
\usepackage{longtable}
\usepackage{changes}
\usepackage{booktabs}
\usepackage[toc, page]{appendix}

\begin{document}

\title{History of Turn-Ons - the irregular 35-day clock - of Hercules X-1}

\author{R.~Staubert\inst{1}, K.~Postnov\inst{2},  V.~Revnivtsev\inst{2,3}, L.~Ducci\inst{1}}

\offprints{R.~Staubert (staubert@astro.uni-tuebingen.de)}

\institute{
	Institut f\"ur Astronomie und Astrophysik, Universit\"at T\"ubingen,
	Sand 1, 72076 T\"ubingen, Germany
\and
	Sternberg Astronomical Institute, Moscow State University, Universitetsky pr. 13, 119234, Moscow,  Russia
	\and 
	Institute of Astronomy of the Russian Academy of Sciences, 48 Pyatnitskaya st., 119017, Moscow, Russia
}

\date{}
\authorrunning{Staubert et al.}
\titlerunning{Irregular 35-day clock in Her~X-1}

\abstract
{Hercules X-1 and Centaurus X-3 are the first ever detected - and as such identified -
binary X-ray pulsars, a ground-breaking achievement by the X-ray satellite UHURU in 1971.
Here we concern ourselves with a particular timing feature, the 35-day flux modulation of Her X-1,
immediately found with its discovery, together with the binary period of 1.7 days and the 
pulsation with a period of 1.24 s.
The super-orbital flux modulation with a period of around 35 days shows two flux maxima, 
a so-called Main-On and a Short-On (with reduced flux). As noticed early on, the period, 
or the length of the cycle, varies, usually by plus / minus half of the binary period.
The system constitutes an irregular clock, with a more or less stable mean period of near 34.85 d.
We consider the repeated rapid increases of  
the flux from an off-state to the Main-On (the so-called \emph{Turn-Ons}) as the best timing marker 
to study this clock. Here we present a complete list of Turn-On times over 55 years, from 
the discovery of Her X-1 to today (2026), based on historic work from the literature
and on observational data provided by the monitoring instruments onboard of five different
X/$\gamma$-ray missions: CGRO/BATSE, RXTE/ASM, Swift/BAT, Fermi/GBM and MAXI. 
We identify to following basic features of the irregular 35-day clock: 
The \emph{Turn-Ons} show preference to occur at orbital phases $\sim 0.25$ and $\sim 0.75$.  
 The duration of the 35-day cycle clusters around three discrete values:
 20.0, 20.5 and 21.0 orbital cycles with a dominant peak at  $20.5 \times P_{\text{orb}}$  ($\sim 57\%$). 
The long-term trend in the 35-day period duration is seen in the O-C (Observed minus Calculated) diagram.
There are two sections, separated by the long \emph{Anomalous Low} of 1999/200 (AL3), which can be 
modeled by parabolic functions with a smooth evolution of the 35-day period. 
 A medium-term variation is observed in the form of \emph{Anomalous Lows}, an 
 essential shut-off (at least of soft) X-rays.  We find an apparent underlying period of ($\sim 5.4\,yr$). 
There are short-term (hundreds of days) variations (in part quasi-periodic) overlaid to 
the long-term trend. Straight "ups" and "downs" in the O-C diagram, demonstrating rather rapid 
$\sim 35\,d$ period changes. 
Finally, we discuss some aspects of the underlying  physics in this accreting neutron star binary pulsar system.}


\maketitle

\vspace{-4mm}
\section{Introduction \label{sec:Introduction}}
\vspace{-1mm}
Hercules~X-1 and Centaurus~X-3 were the first two X-ray pulsars discovered 
by the \textsl{Uhuru} satellite in 1972 \citep{Tananbaum_etal72,Giacconi_etal73}.
Since then both have remained among the most studied X-ray sources.
Her~X-1 is a close binary system consisting of an accreting magnetized 
neutron star and the stellar companion HZ~Her \citep{Liller_72}, 
a main sequence star of spectral type A/F \citep{Crampton_74}. 
The mass of the optical companion is $\sim$2.2$M_{\odot}$  \citep{Deeter_etal81} 
which places the system in between high and low mass X-ray binaries. 
For a more detailed introduction to the Her~X-1/HZ-Her binary system see e.g.
\citet{Staubert_etal09}.
A prominent feature (not discussed here) is  that Her X-1 is the archetype reference
source for Cyclotron Resonant Scattering Features (CRSFs) \citep{Staubert_etal19}.

The X-ray flux of Her~X-1 shows modulations on several different 
time scales: pulsations with 1.24\,s period due to the spin of the neutron 
star (NS), eclipses every 1.7 days with a highly inclined ($\sim 85^\circ$) binary orbit 
\citep{CramptonHutchings_74}, a 35\,d ON/OFF cycle due to variable obscuration of the X-ray source 
by the accretion disk and a 1.62\,d dip cycle (the beat between 1.7\,d and 35\,d),
associated with obscuration by the accretion stream.

\begin{figure}
\includegraphics[angle=-90,width=9cm]{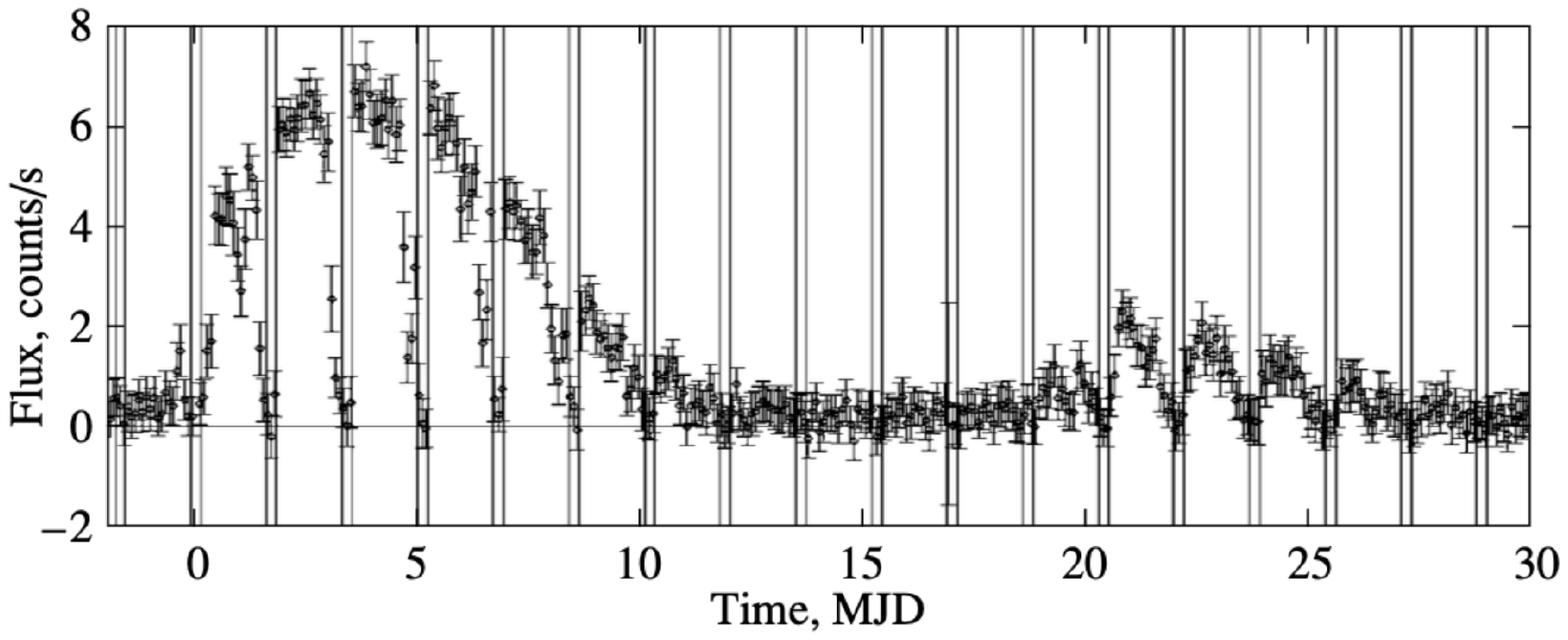}
\vspace{-1.5cm}
\caption{A 35-day light curve as observed by RXTE/ASM
taken from \citet{Klochkov_etal06}.}
\label{fig:35d_lc}
\end{figure}

The 35\,d cycle (Fig.~\ref{fig:35d_lc}) consists of two \emph{ON-states}, separated by two \emph{Off-states}.
The \emph{Main-On} (lasting about seven orbits) is reached after a ($\sim 2d$) increase,
 with a ($\sim 4d$) high flux
interval and a ($\sim 4d$) decline into the first 
\emph{Off-state} (lasting about four orbits). The following  
\emph{Short-On} (lasting about five orbits) has a similar shape as the \emph{Main-On}, 
but only $\sim 1/3$ of its flux. The cycle is completed with a second 
\emph{Off-state} (of about four orbits). So in total $\sim 35\,d$, or 20 to 21 binary orbits.
This shape of the light curve is thought to be due to obscuration of the X-ray source
(the hot spots at the poles of the highly magnetic neutron star) by the tilted and warped 
accretion disk which precesses with a 35\,d period shading the X-ray source. See the 
chapter Discussion for more on the underlying physics.

The  35\,d clock, however, is irregular, the precessional period varies between three
values close to 20.0, 20.5 and  $21.0 \times P_{\text{orb}}$  (34\,d, 34.85\,d and 35.7\,d).
This "quantized" behavior is caused by the fact that the system prefers to choose particular
orbital phases for the \emph{turn-on}, namely around 0.25 and around 0.75. 
This feature had been noted very early on, when Her X-1 was discovered  as
and X-ray pulsar by \emph{UHURU} in 1971  \citep{Tananbaum_etal72},
see also \citet{LevineJernigan_82}.

The irregularity is one key feature we want to concern ourselves in this contribution. We have
followed the 35\,d clock over 55\,yrs - originally motivated by the need to predict the
times of high flux for the planning and scheduling of new observations of Her X-1.
Here, we present a complete list of observed  \emph{turn-on} times and highlight some
of the basic short- and long-term features of the irregular clock in the 35\,d flux modulation 
of Her X-1.

As timing marker for the 35\,d clock we use the times of  \emph{turn-ons}, since
it is evident that the region of the highest flux derivative is best suited for 
timing analysis (we will come back to this point below). The collection of 
\emph{turn-on} times since the discovery of Her X-1 in 1971 is presented in Table \ref{bigtable}.
A representation of the 35\,d clock is given by Fig.~\ref{fig:O-C_all}, the so-called  (O - C) diagram,
showing the evolution of the difference between the observed time (O) (e.g. in MJD) and
the predicted time (C), calculated under the assumption of a constant period.
Here the constant period was chosen to be $20.5 \times P_{\text{orb}}$, that is
$20.5 \times 1.700\,d$ = 34.85\,d (following Staubert et al.\ 1983).
As seen in Fig.~\ref{fig:O-C_all}, the (O - C) values wander off from the mean ephemeris in
an obvious irregular way. An important question is whether this is due to pure
statistical variations (e.g. a "random walk", as suggested by Staubert et al.\ 1983), 
or due to underlying physical reasons.

We like to point out here that there are data collections over 55\,year for other 
characteristic parameters of the Her~X-1/HZ-Her system, like the X-ray flux,
the period of the pulsation (neutron star spin) and the shape of the pulse profiles,
various spectral parameters (power law index, cut-off energy, folding energy),
the energy of the cyclotron line, the optical flux and more. There are correlations
between the different parameters, allowing an insight into underlying physical 
processes. 

The structure of the paper is as follows: 
After the Introduction (Section \ref{sec:Introduction}) to the X-ray binary Her~X-1 and its 35-day
flux modulation, which is the main subject of this contribution, we summarize all the sources to
our database (Section \ref{sec:database}), followed by a description of the informations given 
in the the long table (Section \ref{sec:Table}) - the times of the 35-day \emph{Turn-Ons} and
related values. 
In Section \ref{sec:Characteristics} we discuss special characteristics of the times of the 
\emph{Turn-Ons}, followed by a summary of the long-term structure of the (O - C) diagram
(Section \ref{sec:LongTerm}).
We conclude with Section \ref{sec:Discussion}, a discussion about some of the underlying 
physics in Her~X-1.

\vspace{-1mm}
\section{The data base \label{sec:database}}
\vspace{-1mm}
The one characteristic parameter discussed here, the time of X-ray 
$\emph turn-on$ as compiled  in Table~\ref{bigtable}, comes from different origins. 
We distinguish between two main sources: 
1) The literature as described in Subsection "Historical data", and 
2) Analysis of original data taken by $\emph monitoring$  instruments on
various satellites \emph{CGRO/BATSE, RXTE/ASM, Swift/BAT, Fermi/GBM} 
and \emph{MAXI} on the \emph{Internationl Space Station} (ISS).

\vspace{-3mm}

\subsection{Historical data}

Historical \emph{Turn-On} times were taken from the literature 
- see details below.

\vspace{-3mm}

\subsection{Data from monitoring instruments}

While the historical data provide a rather non-uniform coverage, because 
dedicated pointed observations of Her~X-1 were spars, the situation changed
drastically with the advent of \emph{Monitoring instruments}, allowing to follow
the flux continuously with sufficient sensitivity and time resolution.

\subsubsection{CGRO/BATSE}
In 1991 the $\emph{Compton Gamma Ray oBservatory}$ (CGRO) was launched
with the  $\emph{Burst and Transient Source Experiment}$ (BATSE) onboard
\citep{Harmon_etal02}.
The eight  NaI(Tl) crystal detectors 
operating at 20 keV to 2 MeV, positioned at the eight corners of the payload 
covered the whole sky, mainly to detect gamma-ray bursts, but were also
capable to monitor (and detect)  sufficiently bright sources through observations 
of earth occultations. The data are public and available from the CGRO data 
archive at  NASAs \emph{HEASARC}\footnote{https://heasarc.gsfc.nasa.gov}. 
The listed flux values for Her X-1 are averages over a specified time interval  
(usually of the order of one day). This allowed  the determination of the times 
of $\emph turn-on$ (accurate to about 1/2 of one binary orbit) and measure 
the complete 35\,d light curves. BATSE operated until 1999.\footnote{
https://astro-gdt.readthedocs.io/en/latest/missions/cgro/batse/\\
finders.html\#finding-continuous-batse-data}\\

\vspace{-4mm}
\subsubsection{RXTE/ASM} 
The next monitor instrument, the \emph{All Sky Monitor} (ASM) on the 
\emph{Rossi X-Ray Timing Explorer} (RXTE) (Levine et al. 1996, ApJ 469, L33), 
provided good data from 1996 to May 2011. The three rotating proportional 
counter shadow cameras were sensitive in the 2-12 keV range, and provided intensity 
values in "dwells " of 90 s duration, a significant improvement in sensitivity and
time resolution as monitors.\footnote{http://xte.mit.edu/ASM\_lc.html
(herx1, dwell)}

\subsubsection{BAT on the Neil Gehrels Swift Observatory}
The \emph{Burst Alert Telescope} (BAT) is a coded mask telescope on the
\emph{Neil Gehrels Swift Observatory} (in short Swift) \cite{Barthelmy_etal05}.
The has an effective area of 5200 $cm^2$ with 32500 CadmiumZinkTelluride pixels
sensitive in the energy range 15 to 150 keV. The coded mask with absorbing lead elements 
is 1m above the detector, leading to a fully coded field of view (FOV) of one steradian
(three steradian partially coded), leading to a daily coverage of $\sim 70$ percent of the full sky.
Again mainly designed for the detection and rapid localization of gamma-ray bursts, it
provides a sensitive All Sky Monitor (1 sigma sensitivity about 4 mCrab). The telescope
provides a very good time resolution: typically 10 flux measurements per day for every source
(of sufficient intensity) in the fully coded field of view. The instrument is fully operable since
early 2005 and still (April 2026) working.\footnote{http://swift.gsfc.nasa.gov}.

\subsubsection{MAXI  on the International Space Station} 
The \emph{Monitoring X-ray Instrument} (Maxi) mounted on the 
\emph{International Space Station} (ISS) \citep{Matsuoka_etal09,Mihara_etal22}.
Two slit cameras, one gas proportional counter for 2 - 30 keV and one X-ray
CCD detector for 0.7 - 7 keV, scan the entire sky every orbit of the ISS (91 min).
The sensitivity is 20 mCrab for one orbit and 4.5 mCrab for one day.
Maxi is still operating today. The monitoring data can be retrieved
from an archive in Japan\footnote{http://maxi.riken.jp}. 

\subsubsection{GBM on Fermi} 
The \emph{Fermi Gamma-ray Space Telescope} (formerly the 
\emph{Gamma-ray Large Area Space Telescope} (Glast)),
\citep{ThompsonWilson-Hodge_22}
launched in 2008 carries the instrument \emph{Gamma Ray Monitor} (GBM) 
\citep{Meegan_etal09}.
It consists of 12 NaI(Na) crystal detectors for the  energy range 10 keV to 1 MeV,
placed around the main payload to see the whole sky, and two bismuth germanate 
(BGO) detectors to extend the energy range to 20 MeV. Again, this monitor is 
mainly for the detection of gamma ray bursts.  Because of its
excellent timing capabilities it can also serve as an all sky monitor for
X-ray pulsars. For Her X-1, GBM is the main source for the evolution of the pulse
frequency, it also provides information on the flux (since the determination of
the pulse period requires some minimum flux). The data can be downloaded from a 
NASA  archive.\footnote{https://gammaray.nsstc.nasa.gov/gb,m/science/pulsars.html}

\subsubsection{Dedicated pointed observations} 
Apart from following Her X-1 by \emph{Monitoring Instruments}, a large
number of dedicated pointed observations were carried out over the decades
by essentially all instruments and facilities available. In a number of such
pointings information was derived about the one feature discussed here:
the \emph{Turn-On}, in those cases remarks are made at the corresponding
entries in Table~\ref{bigtable}.

\begin{figure}
\vspace{-1.0cm}
\includegraphics[width=9.5cm,angle=0]{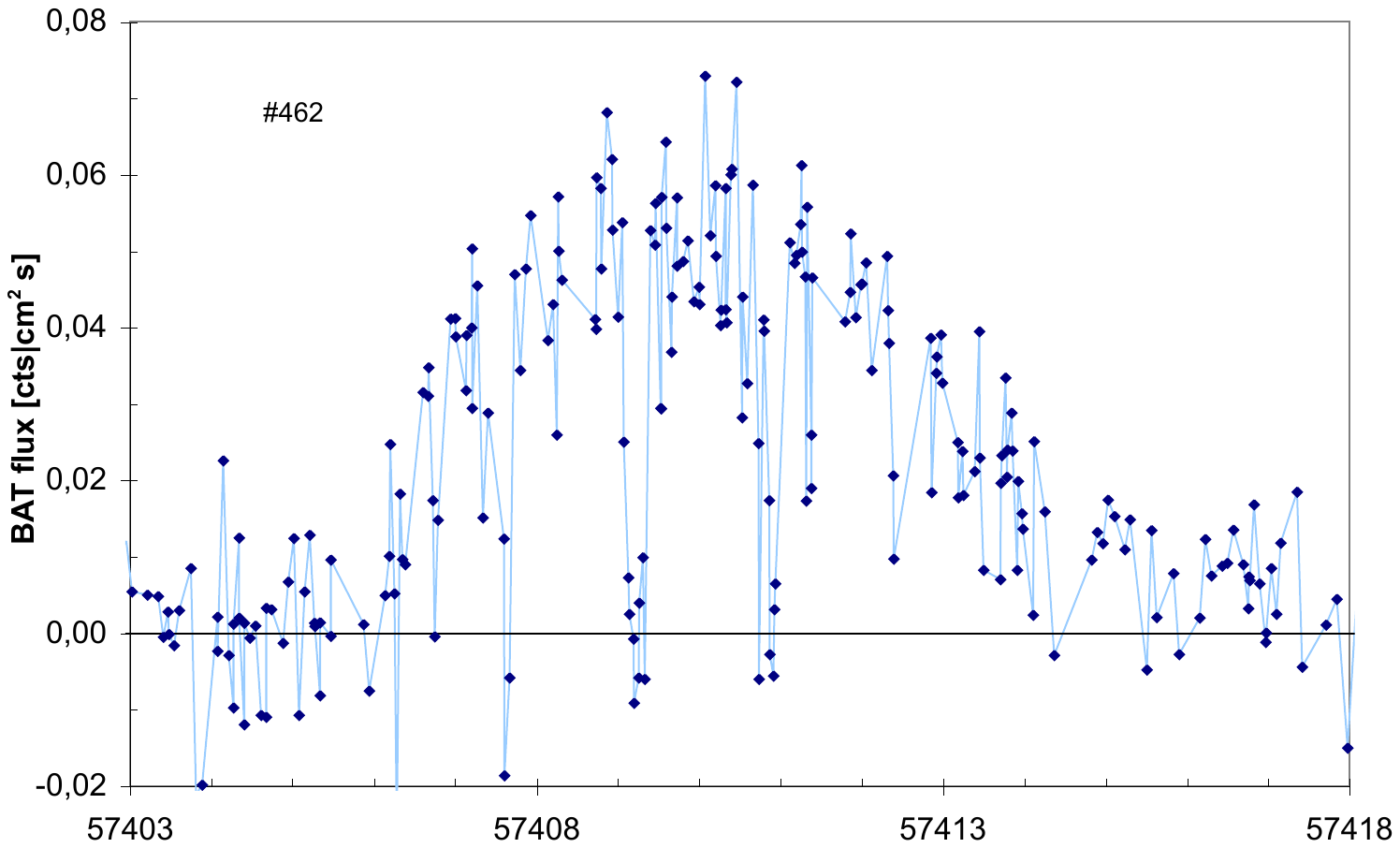}
\includegraphics[width=9.5cm,angle=0]{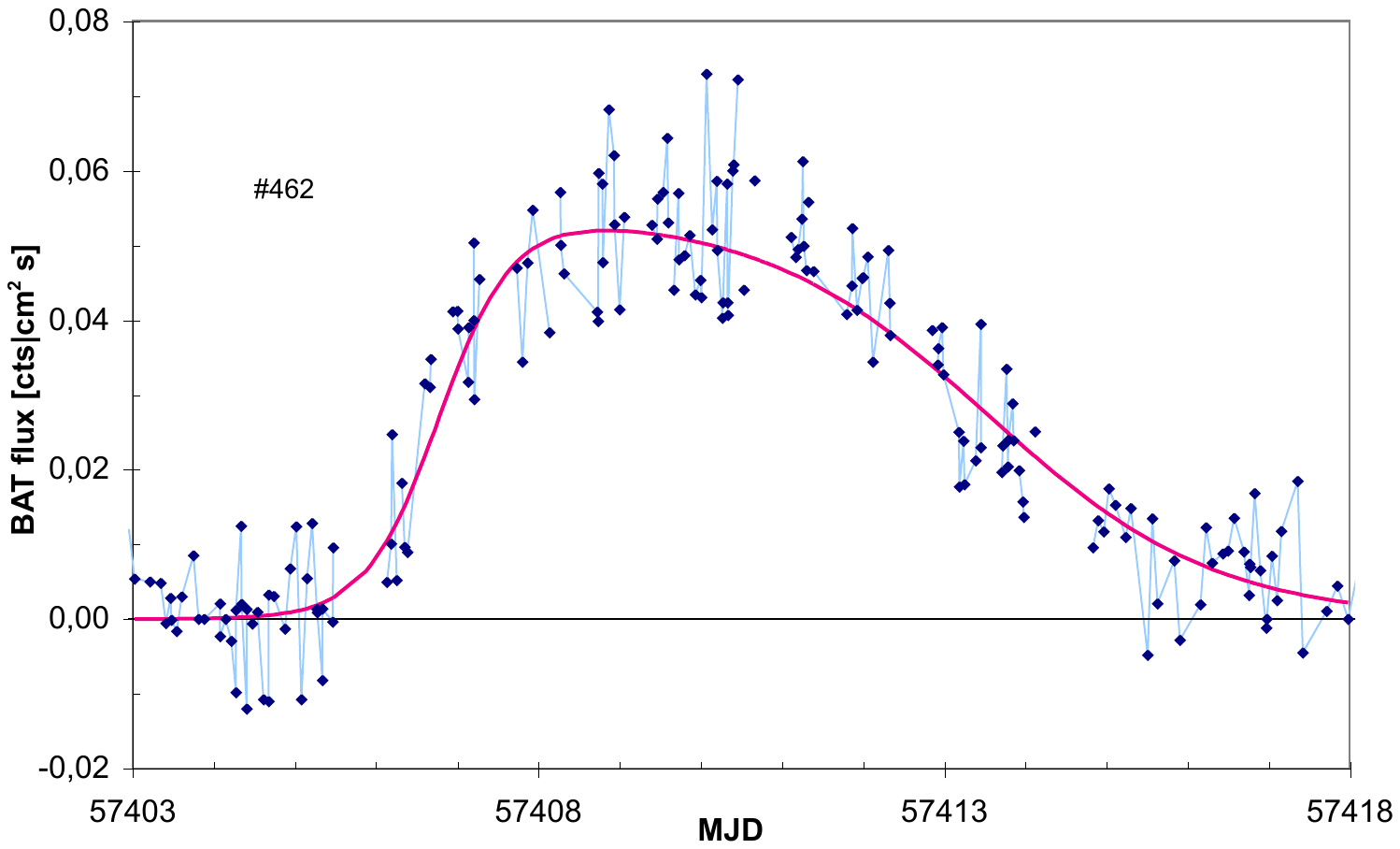}
\vspace{-1.0cm}
\caption{Her~X-1 light curve for the \textsl{Main-On} of 35d cycle 462. The data
points  represent flux averages over one orbit of \textsl{Swift}/BAT. Upper panel:
Lower panel: \textsl{cleaned} data with points during binary eclipses and dips, 
and extreme outliers, removed. Uncertainties of individual measurements around 
the maximum flux are of the order of 5\%. The solid red line is the best fit function 
as defined by equ. (1).
The maximum flux for this 35-day 
cycle is $0.052\pm0.002$ BAT-cts/(cm$^{2}$~s) (taken from Staubert et al. 2016). 
}
 \label{fig:MainOnFit}
\end{figure}

\begin{figure*}
\includegraphics[width=18cm]{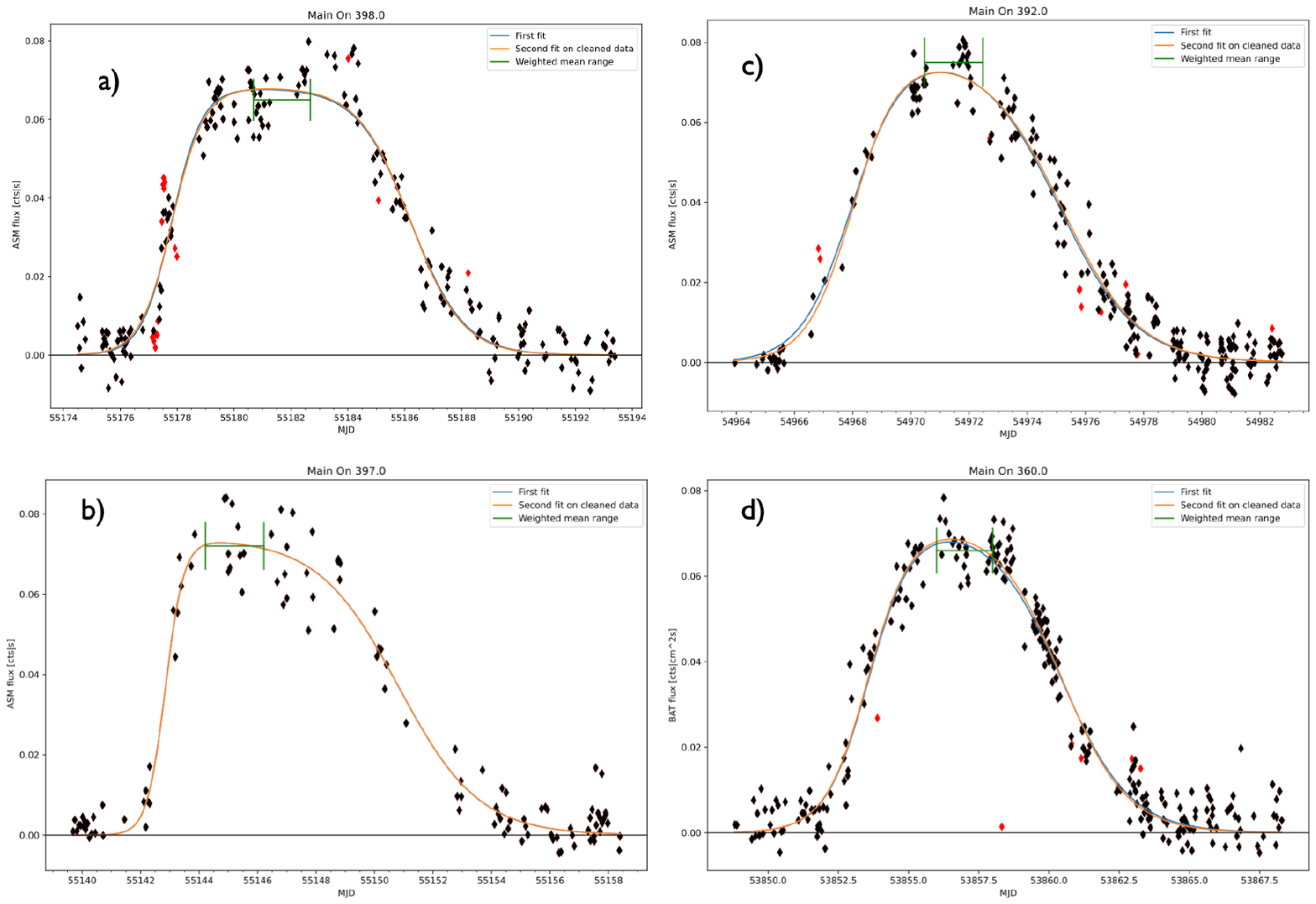}
\caption{Four different Main-On light curves, demonstrating the variability in shape.
Eclipses and dips are removed, the solid curves are from best fits using equ. (1):
the blue curve is the first best fit with automatic removal of data points with deviations 
of > 3 sigma, the red curve is the final fit after removing the data points in red,
found to deviate by more than 2.5 sigma from the first fit (the improvement is generally
not significant).
The maximum fluxes (green horizontal lines) are weighted averages over the range indicated
(note the possible small deviation from the maximal functional value).
a) cycle \#398, with an average rise and a broad "flat top" b) cycle \#397, with a "fast rise"
and medium "flat top"; c) cycle  \#392, "peaking" with a "slow rise" d) cycle \#360, "peaking" 
with an average rise and a fast decay.
}
\label{fig:various}
\end{figure*}

\subsection{Examples of different Main-On light curves }
In this Section we like to present evidence for the above stated fact that the
shape of the \emph{Main-On} light curve (including the \emph{Turn-On})
is quite variable. First, Fig.~\ref{fig:MainOnFit} shows a typical \emph{Main-On} light curve
as observed by \emph{Swift/BAT} (cycle \#462, for the definition of the cycle 
counting see below, e.g., equ. (\ref{equ:O-C}).
The data points are flux averages over one satellite orbit. 
Eclipses and dips, clearly visible in the upper panel, are removed in the lower 
panel. The red line represents the best fit function defined in the text.

Fig.~\ref{fig:various} shows four more observed \emph{Main-On} light curves, demonstrating
the variability of shapes. Particular characteristics are emphazised:
a) "flat-top" (with a rather "fast rise");  b) "slow rise" (with medium "flat top");
c) "peaking"  (with a "slow rise" and "slow decay"); d) "peaking" (with an "average rise" 
and a "fast decay").
As will be shown below, both the time span between the \emph{Turn-On} and
reaching the maximum flux and the duration (width at 30\% of the maximum) vary greatly
(see the wide distributions given in Figs. \ref{fig:start_max} and \ref{fig:width_mainon}).
The width (e.g. duration at 30\% of the maximum) has a fairly wide distribution.

\vspace{-2mm}

\subsection{Definition of  \emph{Turn-On}}
The \emph{Turn-On} is defined by a rapid flux increase 
from a very low level in the OFF-state to a maximum flux in the ON-state (see Figs. 1-3).
Physically, the increase of flux  in the X-ray light curve occurs when the 
tilted, precessing accretion disk moves in such a way to allow a non-obstructed 
view onto the X-ray source itself, which we believe  to be the two small regions
around the magnetic poles of the neutron star (see Section 
\ref{sec:Discussion}).

There is no general convention for its exact formal definition, such that quoted turn-on 
times can only be compared keeping this uncertainty (up to $\sim$0.5\,d) in mind. 
Here we formally define as the time of \emph{Turn-On} (generally given in MJD)
the time where the flux (starting to increase from a background level usually 
very close to zero) reaches 30\% of the maximum flux of the following 
\emph{Main-On}. This time is not significantly dependent on the X-ray energy at 
which the observation had been taken, the vast majority of the data refer  to the 
X-ray energy range of $\sim$3-12\,keV. \\

\vspace{-5mm}

\subsection{How is the  \emph{Turn-On} found?}
Several different sources and methods have been used to find the time of the
\emph{Turn-On} for individual 35-day cycles, as compiled in Table~\ref{bigtable}:
\vspace{-2mm}
\begin{itemize}
\item Historical data (published / compiled).
\item Dedicated pointed observations by the main instruments on various
X-ray satellites.
\item Public original data from monitoring X-ray instruments.
\end{itemize}
The numbering of the 35-day cycles (and \emph{Turn-Ons}) follows that of 
Staubert et al. (1983) (originally proposed by Davison \& Fabian 1977).

\subsubsection{Published 35-day timing data}
We have taken into account published information (proper references are given)
related to 35-day cycles and X-ray \emph{Turn-Ons}: short notes, explicitly stated 
times, indirect determinations through optical observations \citep{Thomas_etal83},
and others. Sometimes the turn-on time was read off published light curves 
(when sufficiently accurate informations were given). These entries are  marked 
by "*" in Table~\ref{bigtable}. In a few cases quoted times were corrected 
(marked by "+" in Table~\ref{bigtable}) when obvious mistakes in referencing 
time had been identified (see comments in Table~\ref{bigtable}).
Otherwise, historical data have been taken from published Tables:
\citet{DavisonFabian_77, Staubert_etal83,
CrosaBoynton_80,Oegelman_87,Baykal_etal93,
ScottLeahy_99,Shakura_etal98,LeahyIgna_10,
Staubert_etal13}. \\

\vspace{-5mm}

\subsubsection{Own dedicated pointed observations}
We have been active in writing observing proposals to various X-ray missions
and used the data of the main instruments of those observations, 
mainly \emph{RXTE}, \emph{NuSTAR}, \emph{INTEGRAL},
\emph{Suzaku}, \emph{SRG/ART-XC} and \emph{HXMT}/Insight).

\subsubsection{Public original data from monitoring X-ray instruments}
The missions providing original X-ray light curves (with different time resolution)
have been listed above. To find the time of  \emph{Turn-On} (either from dedicated 
pointed observations or from monitoring instruments), our initial way to find the 
ime of  \emph{Turn-On} was to plot the light curve (usually in units of counts/s), 
then identify the peak flux and the 30\% level of the same and read the time at 
which this level is reached "by eye". 

Later "objective" methods were applied. These
include smoothing the data (to reduce the statistical scatter) 
and formal best fits (using minimum Chi-squared) with various heuristic 
functions. We describe two such functions below: \\ 
- First, a linear fit (a straight line) to the rapidly increasing flux values,
either by defining the time interval of the used data "by hand" or by 
introducing formal rules to exclude data (around the start and the stop)
with larger deviations from such a straight line. Again, the time of 
\emph{Turn-On} is then the time where the flux reaches 30\% of the maximum 
flux of the following \emph{Main-On}. \\
- Second, a formal fit to the complete \emph{Main-On} light curve, including
sufficient data (one to two days) of the low-state before the turn-on and
sufficient data of the decay into the low-state (close to zero flux) with the
following function (see \citet{Staubert_etal16}):

\vspace{-3mm}
\begin{equation}  
{\rm F}(\rm t) = A_0 \times \frac{1}{1+exp(\frac{A_1-t}{A_2})} \times \frac{1}{1+exp(\frac{t-A_3}{A_4})}   
\label{equ:main-on-fit}
\end{equation}

Of course, data of eclipses and dips are removed, see e.g. Fig.~\ref{fig:MainOnFit}.
The solid red line is the best fit to the cleaned light curve of 35d cycle \#462. 
Again, the time of  \emph{Turn-On} is  the time where the flux reaches 30\% of 
the maximum flux.
The latter method is greatly preferred and offers additional benefits, by
providing further characteristic values, like the maximum flux, the time of the
flux falling below the 30\% (allowing to measure the duration of the 
\emph{Main-On}), as well as the steepness of the increase and of the decay
of the flux. Corresponding distributions of those parameters are shown below. 

Another method is to make use of the fact that the 1.24-sec pulse 
profiles show a strong and reproducible evolution with 35-day phase:
\citet{Staubert_etal13} have developed a technique of "pulse profile fitting",
showing that the two clocks, the "turn-on clock" and the "pulse profile clock" are 
synchronized. This has recently (March 2026) turned out to be quite useful: during the
latest Anomalous Low the monitoring instruments \emph{Swift/BAT} and \emph{MAXI} 
were not able to see the source (at low X-ray energies), but pulse profiles 
in the 9-13 keV range and 3-30 keV range were well observed by 
\emph{NuSTAR}, and the turn-on times could be extrapolated from
multiple \emph{Main-On}  observations, using  pulse profile templates.
Applying this method to \emph{NuSTAR} data of February-April 2026,
it was found that the 35-day period tended to be short (18 to 19.5 times the 
orbital period), as had also been observed in the long \emph{Anomalous Low}
(AL3) in 1999/2000 \citep{Staubert_etal09}.
This allowed a more accurate prediction of the time of occurance
of the maximum flux in future 35-day cycles (used  in planning further observations).

\begin{figure}
\includegraphics[width=10.5cm]{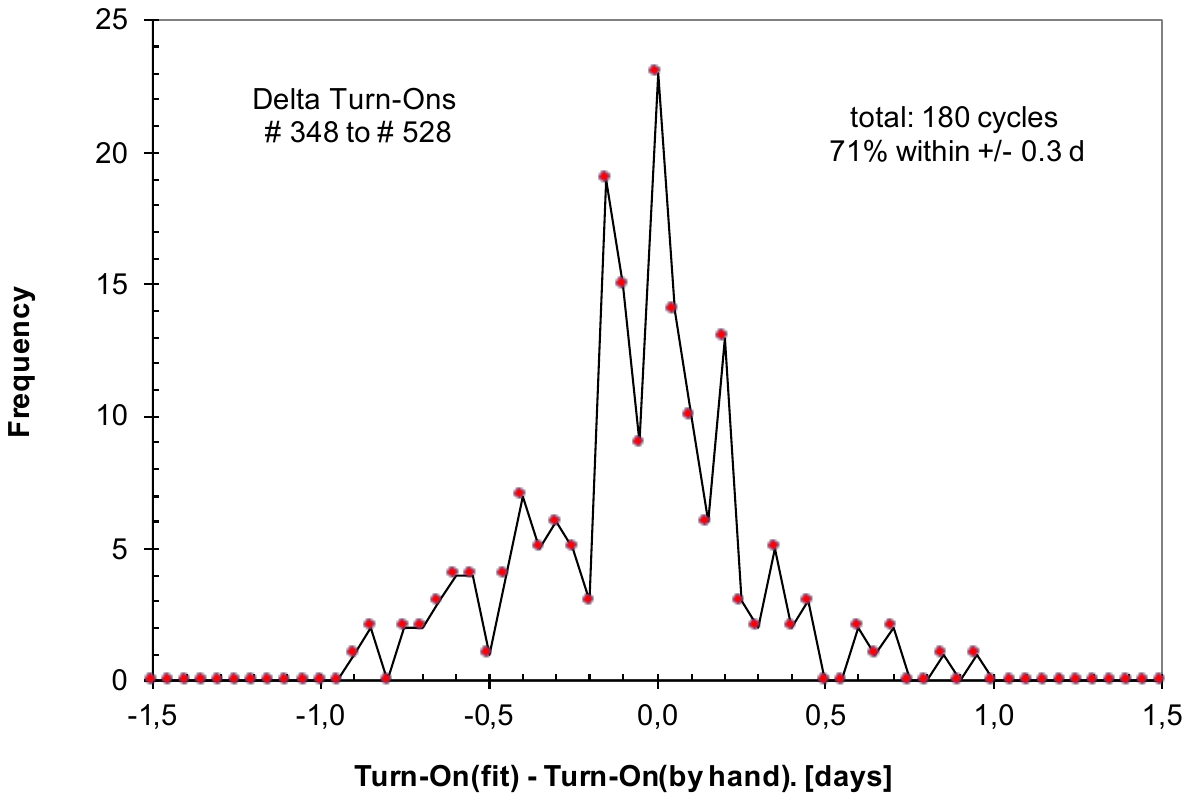}
\vspace{-13mm}
\caption{Comparison of finding the turn-on time "by hand" (a trained eye) 
versus the formal fitting of the Main-On flux - for a sub-sample of data from
Swift/BAT. }
\label{fig:compare}
\end{figure}

In cases when the original data are sparse (or corrupted)
by instrumental or other observational difficulties (usually recognizable by large
uncertainties, e.g. 0.5\,d or larger), we have applied rules of systematic behavior
the Her X-1 system is following (e.g. the preferred turn-on phases or the length of
cycles, see below). So the stated turn-on times are probably physically more
accurate than the listed uncertainties suggests.

\subsubsection{Methods, not to be used}

It has been attempted, e.g.  by  \citet{LeahyIgna_10}, to find the \emph{Turn-On}
time by using the time of the maximum flux in the \emph{Main-On}, 
through template fitting of the entire 35-day light curve, and then extrapolating
backwards, assuming a constant time delay between the \emph{Turn-On} and the
time of the maximum flux. This procedure suffers from three problems: first, when 
it comes to timing, it is the best to select regions where there is the highest flux 
derivative, not the flat parts of the peaks. Clearly, this is the \emph{Turn-On} itself, 
not the region around the peak of the Main-On; second, the shape of individual
Main-On light curves varies, the region around the maximum is quite broad
(see Fig.~\ref{fig:various}); and third, the values of the time 
difference between the turn-on and the maximum flux follow a rather wide
distribution (see Fig.~\ref{fig:start_max}). So, completely wrong conclusions 
have been reached (e.g. the claim of a uniform distribution of the 
\emph{Turn-Ons}  in binary phase).

\begin{figure*}
\includegraphics[width=18cm]{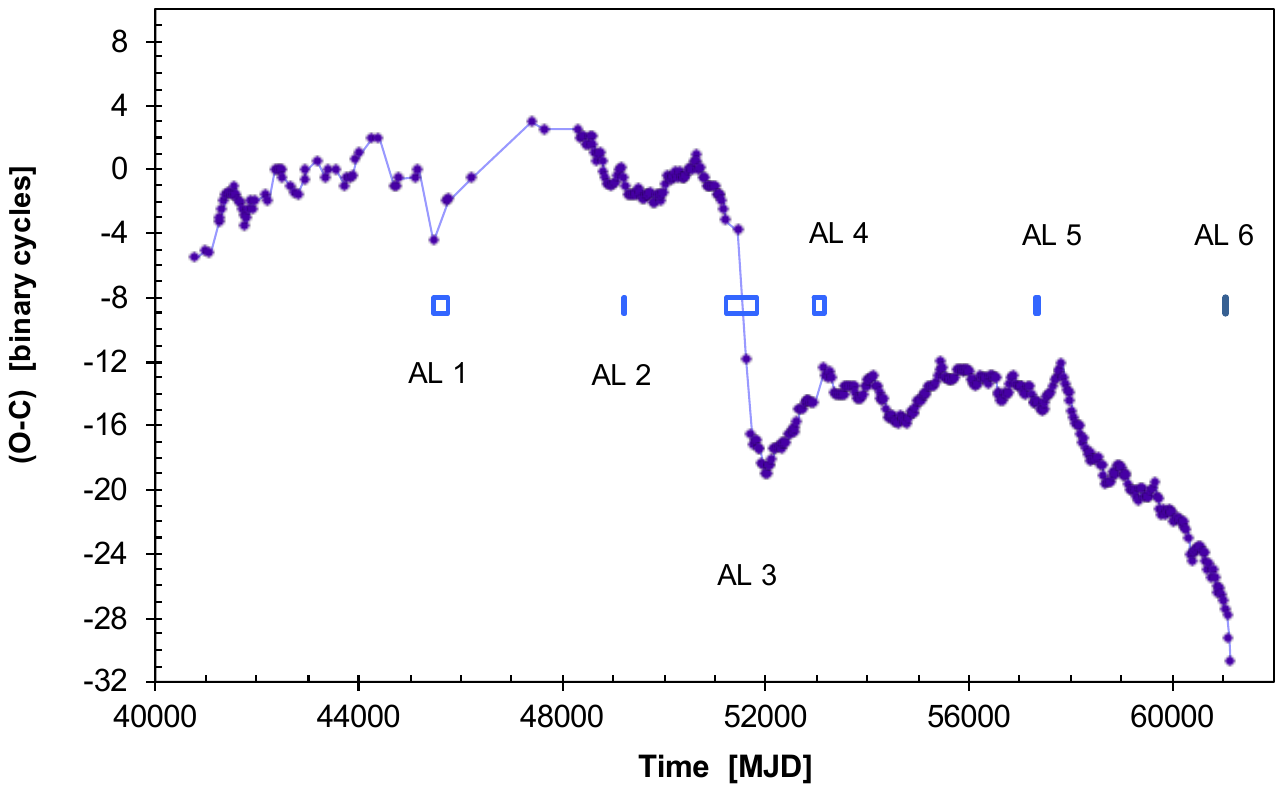}
\vspace{-1.5cm}
\caption{The (O - C)-diagram: the long-term (1971--2026) global structure 
of the 35-day clock in Her X-1. O-C is the difference between observed (O) 
and calculated (C) Turn-On times. The calculated time follows the epoch 
of Staubert et al. 1983 \citet{Staubert_etal83}, using cycle no. 31 as a 
reference. (O-C) = zero for the 
31$^{\rm st}$ turn-on, observed at $T_{31}$ = MJD 42409.943.
Other Turn-On times are calculated using a constant 35\,d period
of 20.5 $P_\text{orbit}$ with $P_\text{orbit} = 1.700167788$ d.
The open rectangles mark the times and durations of observed 
"Anomalous Lows".
}
\label{fig:O-C_all}
\end{figure*}

\begin{figure}
\includegraphics[width=10.0cm]{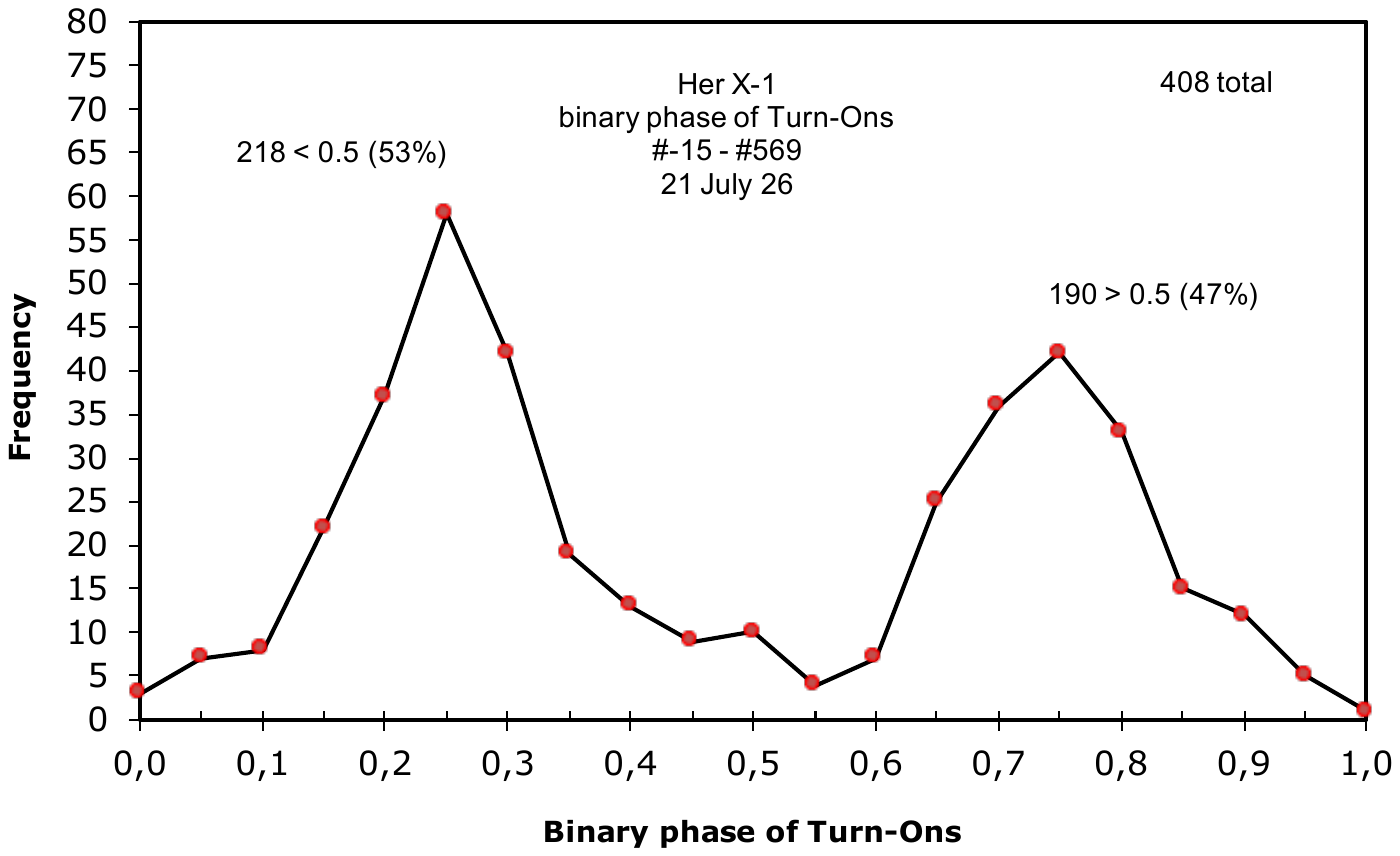}
\vspace{-13mm}
\caption{Frequency distribution of binary phases at which the turn-on occures,
demonstrating that turn-ons prefer to happen around phase 0.25 and 0.75. }
\label{fig:bin_phase}
\end{figure}

\begin{figure}
\includegraphics[width=11cm]{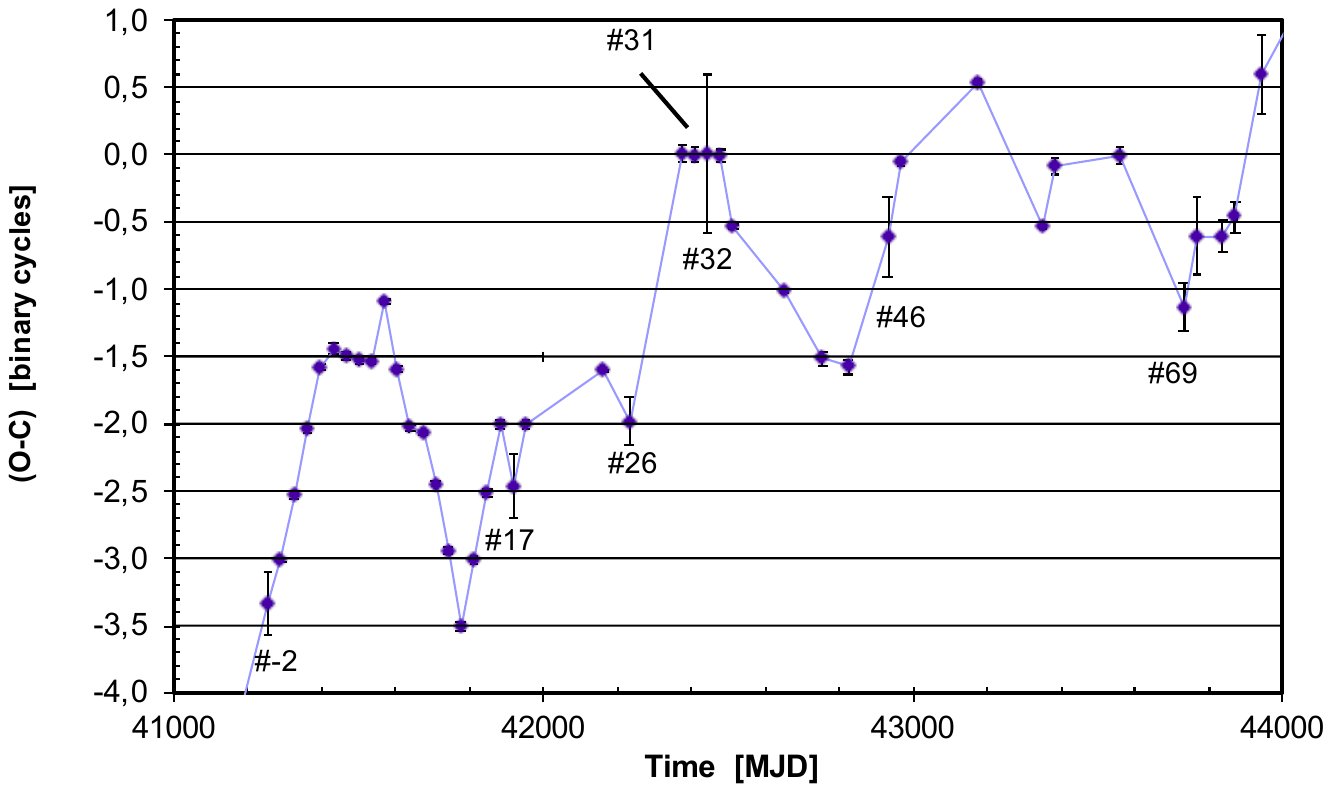}
\vspace{-16mm}
\caption{O-C diagram from the early times of Her X-1 observations
(\#-7 to \#2, roughly 1971-1978), showing that (O-C) is a quantized parameter
with a spacing of 0.5 $P_{orb}$, a consequence of the preferred turn-ons
around 0.25 and 0.75 in orbital phase. The early data (from UHURU) also
show that changes of (O-C) seems to only happen by steps of 0 or 
+/- 0.5 $P_{orb}$, a rule that is now firmly established.  
Going up one step corresponds to a cycle length of 21 $P_{orb}$,
going down one step means  20  $P_{orb}$, staying on the line means
20.5 $P_{orb}$. 
}
\label{fig:O-C_start}
\end{figure}

\section{Table of \emph{Turn-Ons} - the (O~-C) diagram \label{sec:Table}}
The \emph{Turn-On}  values listed in Table~\ref{bigtable} represent the "best" values,
often arrived at by combining different methods. We like to note that in
producing this final table, tests were performed to control the overall
consistency between the different methods. It was found that the times 
determined by a "well trained eye" were generally quite consistent  with
those from formal fits (see Fig.~\ref{fig:compare}).
This result gives us confidence that the uncertainties
quoted are realistic (even for those data from the early days when no
archived original data were available). \\

Table~\ref{bigtable} gives the following information:  \\
- Col.~1: cycle number (following  Staubert et al. 1983).\\
- Col.~2: time of \emph{Turn-On} in MJD (exact definition see above). \\
- Col.~3: estimated uncertainty of \emph{Turn-On} in days. \\
   For the early data (\#-15-\#140) the uncertainties from the cited references
   have been copied (likely some are under-, others over-estimated). \\
 -Col.~4: Number of binary cycle just before the \emph{Turn-On} : the center of eclipse 
  of binary cycle 0 is at MJD 46359,87194, P(orb) = 1,70016759\,d, dP(orb)/dt = -4,85E-11\,s/s
  \citep{Staubert_etal09} . \\
 - Col.~5: Orbital phase of \emph{Turn-On}, using the ephemerneris of  \citet{Staubert_etal09}. \\
 - Col.~6: Length of 35-day cycle (the difference between adjacent \emph{Turn-Ons}, 
   in case of gaps these values are an "average length"). \\
- Col.~7: (O - C), see  equ. (\ref{equ:O-C}):
the difference between observed ("O") and calculated / expected  
\emph{Turn-Ons}, "C" is found using the reference (O - C) = 0 at cycle \#31 (MJD 42409.9) 
   and a constant cycle length of 20.5 x P(orb) = 34,8534356\,d (see Fig.~\ref{fig:O-C_all}). \\
  - Col.~8: Satellite or instrument as source of observational data. \\
  - Col.~9: Reference (see end of Table~\ref{bigtable}). \\
  
 The key graph of this contribution - the \emph{(O - C) diagram} - is given in Fig.~\ref{fig:O-C_all}. 
 It is a representation of the 35-day flux modulation of Her X-1, demonstrating
 the short- and long-term variability of this clock. Below we will discuss various
 characteristic features of this clock.

\section{Characteristics of the 35-day clock \label{sec:Characteristics}} 
Here we like to summarize characteristica of the 35-day modulation
of the X-ray flux of Her X-1. The first obvious and fundamental feature is that
the 35-day clock is irregular - which was noted early on by analyzing the very
first X-ray observations of the source by \emph{UHURU} \citep{Tananbaum_etal72}.
Fig.~\ref{fig:O-C_all} demonstrates the behavior of the 35-day clock over the complete 54 years
over which Her X-1 has been known and observed as an X-ray source. Displayed is the so-called
the (O - C)-diagram, that is the difference between observed (O) and calculated (C) 
\emph{Turn-On} times. 
The calculated times are from the epoch of \citet{Staubert_etal83}, 
using cycle no. 31 as a reference: (O-C) = zero for the 31$^{\rm st}$ turn-on, 
observed at $T_{31}$ = MJD 42409.943.
The calculated turn-on times are found by assuming a constant 35\,d period
of 20.5 x $P_\text{orbit}$ with $P_\text{orbit} = 1.700167788$ d:

\vspace{-5mm}
\begin{equation}
T_\mathrm{MJD} = 42409.943 + 20.5 \times P_\mathrm{orb} \times (N_{35} - 31).
\label{equ:O-C}
\end{equation}

For a strict clock an (O-C)-diagram would display a straight line, ideally a constant
value if the period (and its derivative) is exactly known. Obviously, this is not 
so for Her X-1. Immediately recognizable are a long-term evolution: two epochs in time, 
both somewhat longer than 20 years, each with an apparent overall parabolic behavior,
separated by a dramatic discontinuity. This discontinuity is coincident with an event
observed in 1999 / 2000, when the X-rays were basically shut-off for a time of about
602 days. This event is called an \emph{Anomalous Low} (AL), actually the third and
longest such event out of a series of six (up to now) events of similar nature (see 
Section \ref{sec:AL} on Anomalous Lows below). We will also briefly come back to the 
long-term parabolic evolution (Section \ref{sec:LongTerm}), which is the main subject of 
\citet{Revnivtsev_etal26}.
The second obvious feature is
a clear variability on time scales of years (see Fig.~\ref{fig:O-C_start}) - in the early days 
suspected to be possibly of pure statistical nature, e.g. a \emph{Random Walk} 
\citep{Staubert_etal83}. We will come back to this aspect below.

\begin{figure}
\includegraphics[width=10.0cm]{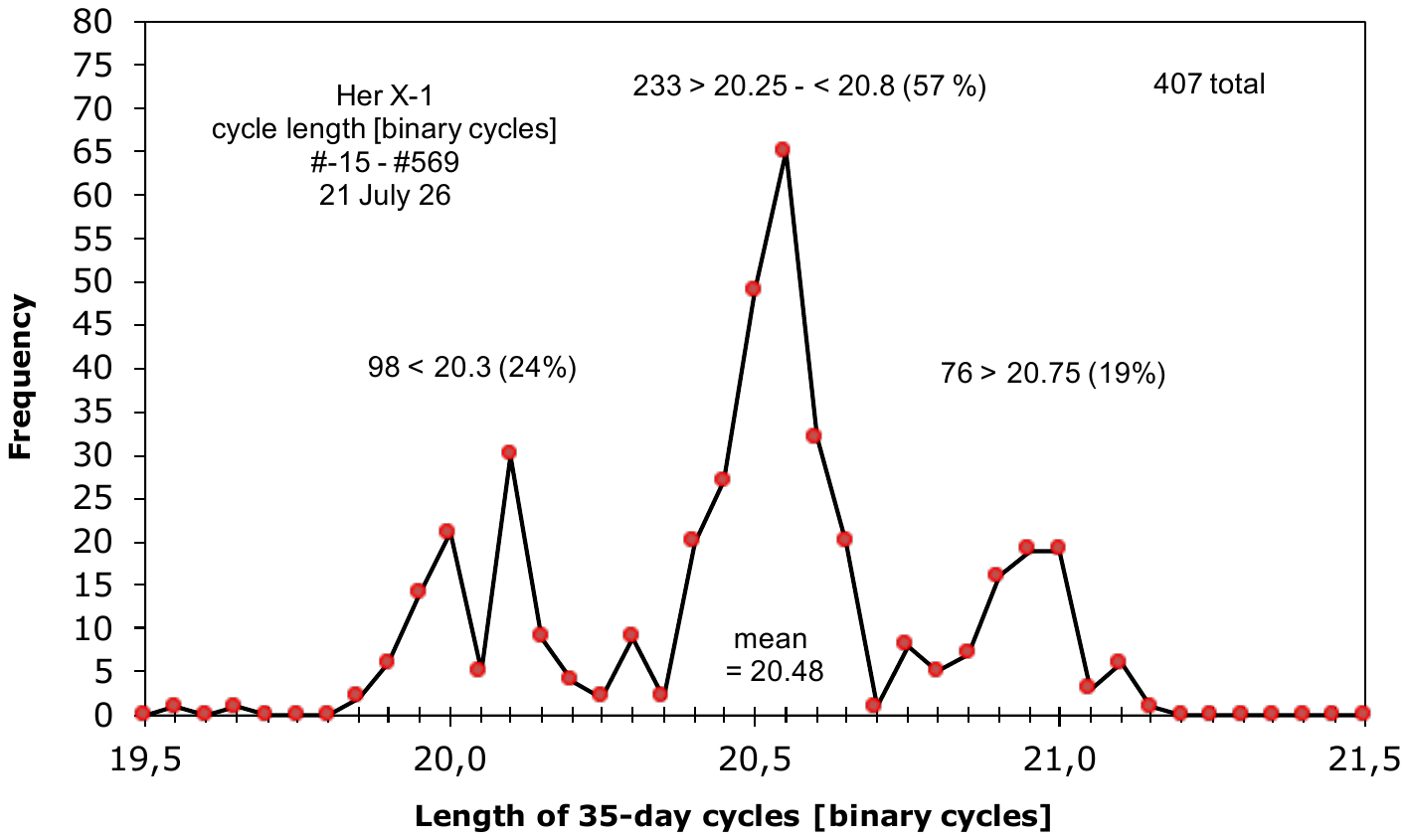}
\vspace{-10mm}
\caption{Frequency distribution of the length of the 35-day cycle in units of 
binary cycles (1.700 d). The length of 20.5 x $P_{orb}$  (35.85\,d) is most abundant. }
\label{fig:length}
\end{figure}

In Fig.~\ref{fig:O-C_start} the short-term variability is demonstrated by data from the first $\sim$10 years
of observing Her X-1.  Apart from the (semi-systematic) ups and downs it is clearly seen
that the data points are on (or very close to) horizontal lines, that are locations of constant
(O-C), representing a constant 35-day period of 20.5 $P_\text{orbit}$. In addition, it is
evident that from one cycle to the next the (O-C)-value changes by discrete steps of 
half an orbital period: one up, no step, or one down. This behavior can be understood
when one looks at the orbital phase at which the \emph{Turn-Ons} actually happen:
the system prefers to turn on near orbital phases 0.25 and 0.75, which was found 
(already quite early from the first observations by \emph{UHURU}, see \citet{LevineJernigan_82}). 
This is demonstrated in Fig.~\ref{fig:bin_phase}, frequency diagram of binary phases of \emph{Turn-Ons}.
(containing all 410 values (until \#570) of orbital phase values listed in Table~\ref{bigtable}.

\begin{figure}
\includegraphics[width=10cm]{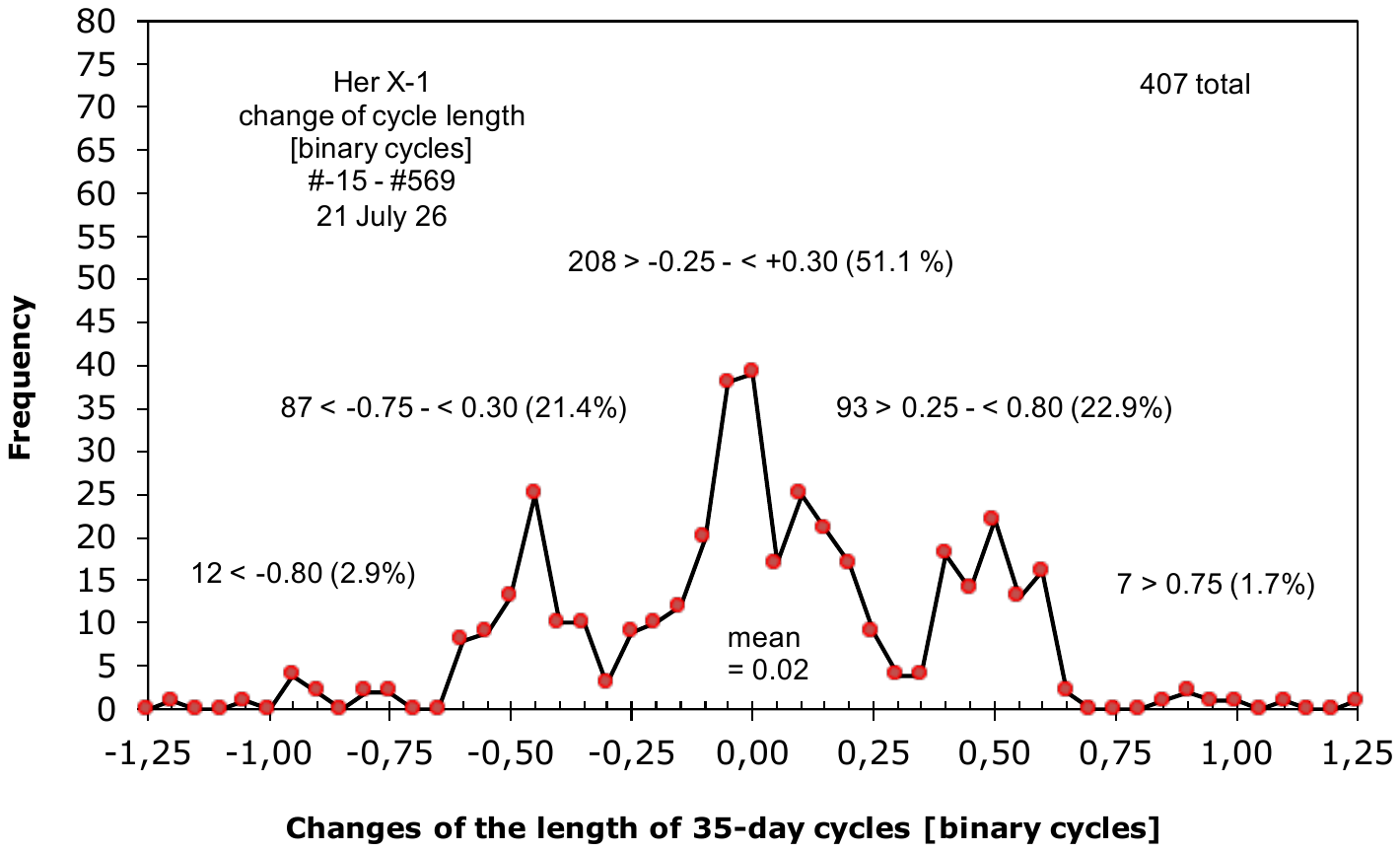}
\vspace{-10mm}
\caption{Frequency distribution of the change (from cycle to cycle) in length of the
35-day cycle (in units of the binary cycle = 1.700\,d).}
\label{fig:change_length}
\end{figure}

In summarizing the basic behavior of the system with regard to the
times of its X-ray \emph{Turn-Ons} by stating  that the Her X-1 / HZ~Her binary system 
obeys the following rules throughout its entire observed life (54 years) - with a few, 
but notable exceptions (to be discussed below): \\
- X-ray \emph{Turn-Ons} happen preferentially around orbital phases 0.25 and 
0.75 (Fig.~\ref{fig:bin_phase}) (albeit with a typical uncertainty of the order of +/-~0.1). \\
As a consequence the following rules are also true: \\
- Rule 1: The data points in the (O-C)-digram are "quantized" (Fig.~\ref{fig:O-C_start}), they are on 
or near horizontal lines, half an orbital cycle apart; from one 35-day cycle to the next the 
movement is along one such line, one up or one down. \\
- Rule 2: The length of an individual 35-day cycle (defined as the time from one \emph{Turn-On}
to the next) is generally either 20.0, 20.5, or 21.0 x $P_\text{orbit}$ (Fig.~\ref{fig:length}).
There is, however, some scattering around these mean values, meaning that the quantization 
(of Rule 1) is blurred and data points of (O-C) can end up in-between of those horizontal lines. \\
- Rule 3: The change of the length from one cycle to the next is generally either zero, or plus or
minus 0.5 x $P_\text{orbit}$ (Fig.~\ref{fig:change_length}, again with a degree of statistical blurring).
However, there are physically significant exceptions with changes near plus or minus
1 x $P_\text{orbit}$ (see the discussion of exceptions to the rules below). 

Fig.~\ref{fig:length} shows that the dominant cycle length is around 20.5 x $P_\text{orbit}$:
57\% of all cycles listed in Table~\ref{bigtable}.

\begin{figure}
\includegraphics[width=10cm]{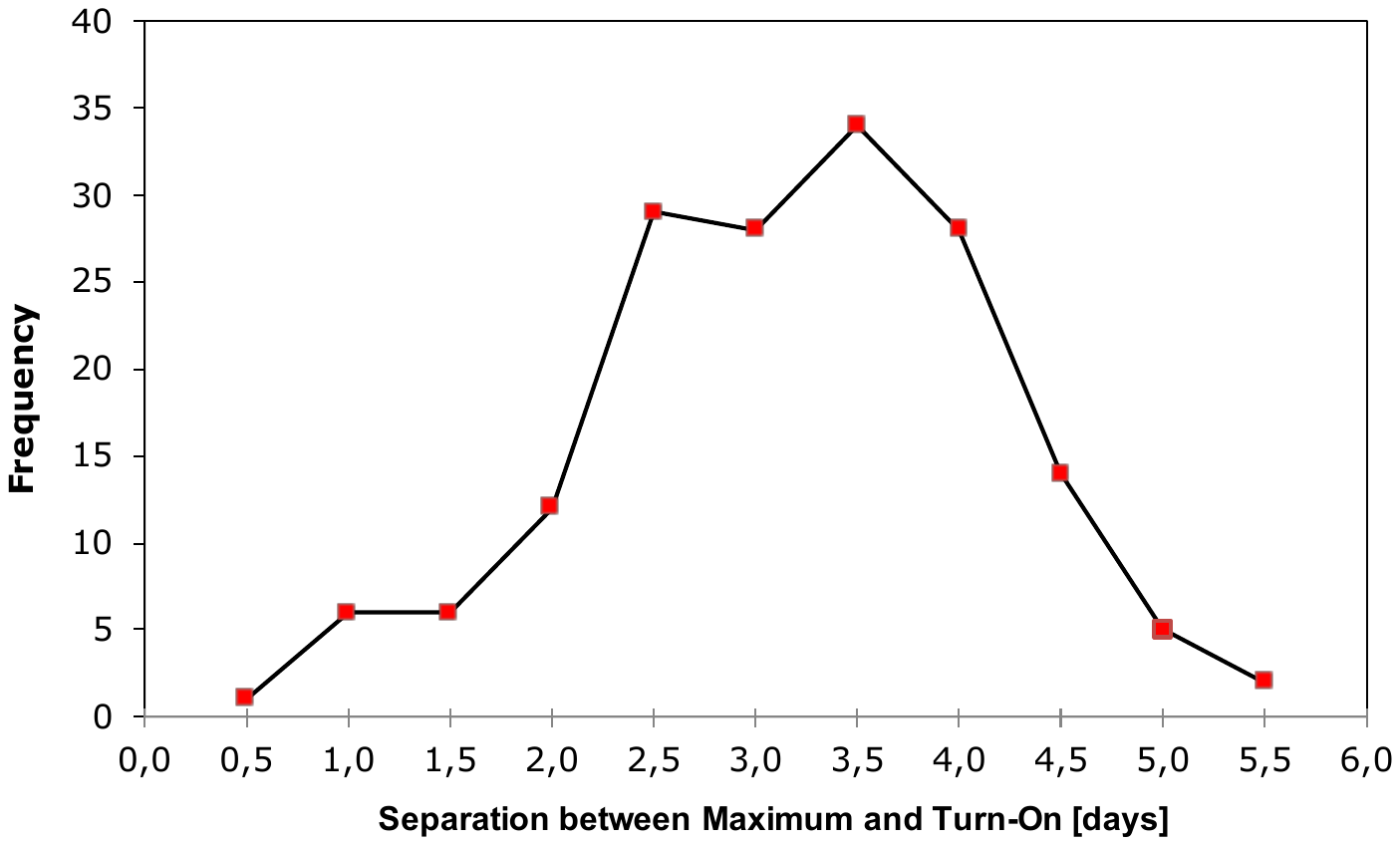}
\vspace{-12mm}
\caption{Frequency distribution of the time span from \emph{Turn-On} to the
maximum flux, as determined by formal fits  with equ. (1) of 163 \emph{Main-On} 
light curves observed by BAT/\emph{Swift}. The average value is near 3.0 days.
Note the wide range from below one to about three binary orbits.}
\label{fig:start_max}
\end{figure}

\begin{figure}
\includegraphics[width=10cm]{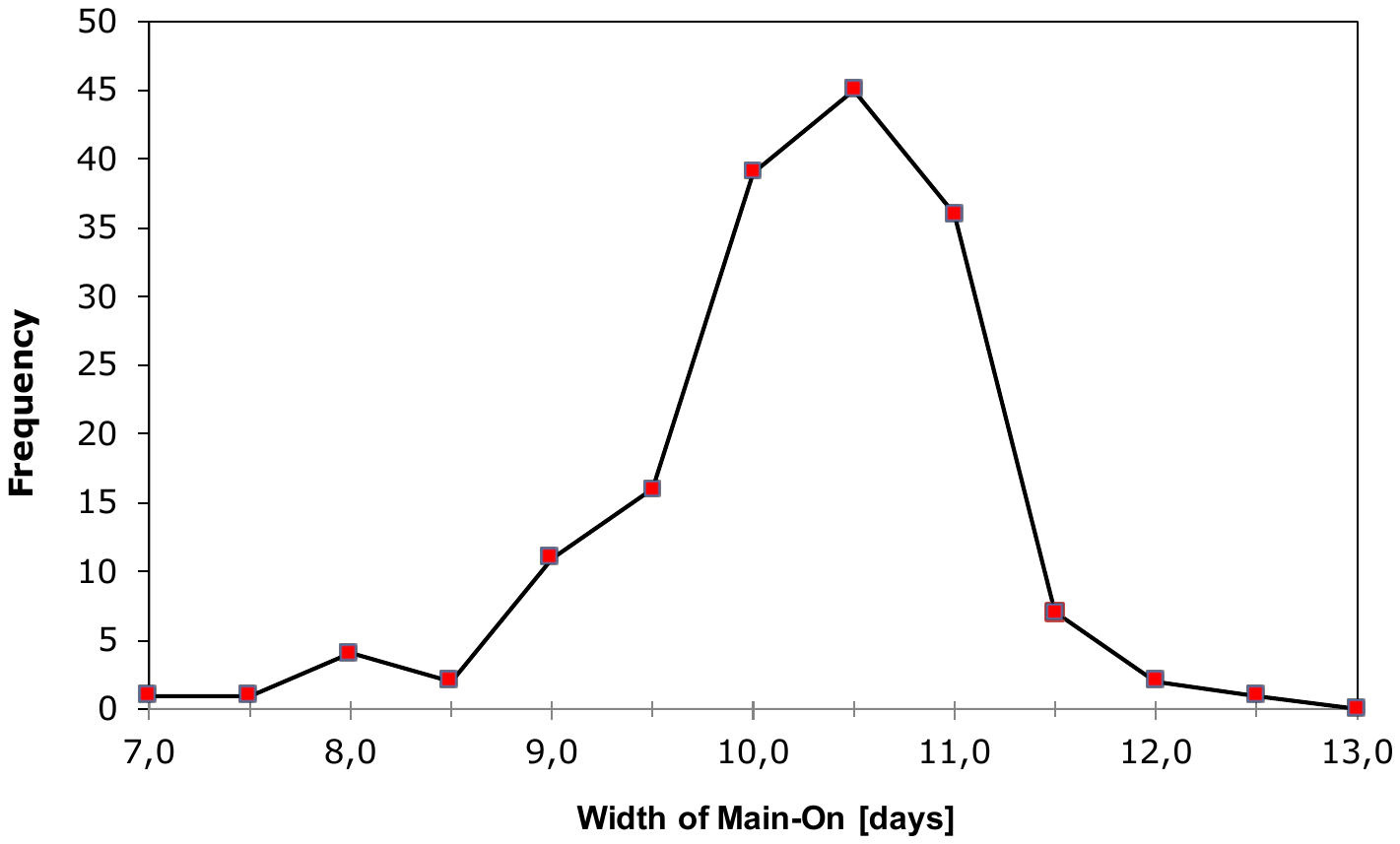}
\vspace{-12mm}
\caption{Frequency distribution of the width of the \emph{Main-On} 
(width at flux above 33\% of the maximum), from formal fits  with equ. (1) of 163 \emph{Main-On } 
light curves observed by BAT/\emph{Swift}. The average value is 10.0 days.}
\label{fig:width_mainon}
\end{figure}

Two further parameters are of interest: the separation of the time of
the maximum flux in the \emph{Main-On} and the \emph{Turn-On}
(Fig.~\ref{fig:start_max}) and the width of the \emph{Main-On} itself -
from reaching 33\% of the maximium flux to falling below 33\% 
(Fig.~\ref{fig:width_mainon}) (found for the sub-set of those cycles for
which a best fit of the \emph{Main-On} with the function of equ. (1)
had been done, that is BAT/\emph{Swift} cyles \#349 to \#528). 
So, the maximum flux in the \emph{Main-On} is typically reached about
two binary cycles after the \emph{Turn-On}, and the duration  of the
\emph{Main-On} (above 33\% of the maximum flux) is mostly
between 9.5 and 11.5\,days. Both distributions are rather wide and
explain, why the attempt to find the time of \emph{Turn-On} by
measuring the time of the maximum flux and subtracting a constant value
must lead to the completely wrong conclusion reached by 
\citet{LeahyIgna_10} and \citet{LeahyWang_20} of a uniform distribution of
orbital phases of \emph{Turn-Ons}. It is clear that the combined variability 
of the two (un-correlated) parameters destroys any structure in turn-on phases.

\begin{figure}
\includegraphics[width=10.5cm]{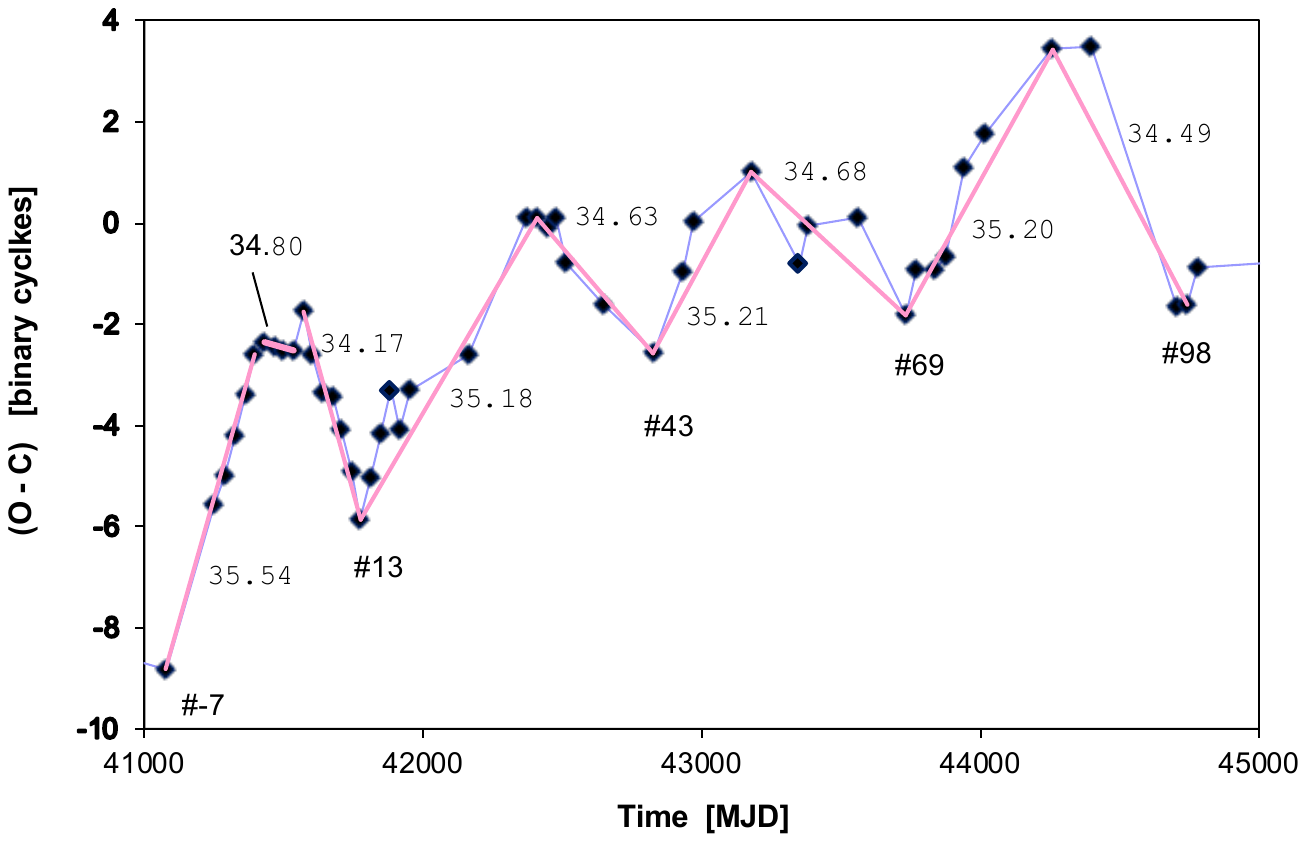}
\vspace{-8mm}
\caption{O-C diagram for early 107 precession cycles (roughly 1971-1978), 
showing fine structure, with groups of "ups" and "downs" (the mean cycle lengths
in units of days for those groups are given). The 35-day cycle length can change
on very short time scales from  21 $P_{orb}$ ("up") to 20 $P_{orb}$ ("down") - in
a quasi-periodic way (here with a period of $\sim$ 900 days).
See a more detailed analysis of the fine structure of the (O-C) diagram for the
complete data set of 55 years \citep{Revnivtsev_etal26}.
}
\label{fig:O-C_start2}
\end{figure}

\subsection{Exceptions to the rules}
Apart from the statistical blurring of the above discussed rules
about the lengths of cycles and its changes from cycle to cycle
(obvious from the widths of the distributions in Fig.~\ref{fig:length} and 
Fig.~\ref{fig:change_length}), there are a few significant exceptions to 
those rules: \\
- Rule 1: From Table~\ref{bigtable} we find four entries with cycle length
< 19.75 x $P_\text{orbit}$: cycle Nos. 291/296, apparently associated with 
AL3 and Nos. 568/569, associated with AL6 (see the Section 
\textsl{Anomalous Lows} below). In addition, two further cycles: No. 545
and No. 558, with no apparent association. 
There are no entries with length > 20.5 x $P_\text{orbit}$.\\
- Rule 2:  Fig.~\ref{fig:length} demonstrates the scatter around the 
preferred cycle lengths of either 20.0 / 20.5 / 21.0  x $P_\text{orbit}$. \\
- Rule 3:  Fig.~\ref{fig:change_length} shows that the rule not to change 
the length of cycles (from cycle to cycle) by a full binary
cycle is violated more often: because of the statistical scatter we
define the condition to plus or minus 0.75 x $P_\text{orbit}$: 
this happens in seven (+) and 12 (-) cases, together $\sim 5\%$ 
of all entries in Table~\ref{bigtable}. \\
We regard all the above listed "violations" of the rules as physically significant,
in all cases there is no room of enforcing the rules without greatly miss-interpreting
the given observational data. We conclude that the dynamics in the
binary system Her~X-1 / HZ~Her can produce - at times - deviations from otherwise
general behavior. \\

\vspace{-5mm}
\section{Variability properties of the 35-day clock \label{sec:LongTerm}}
The timing properties of the variability of the 35-day clock of Her X-1 can be 
categorized into four typical time scales: \\
- "Cycle-to-cycle": 35\,days,  \\
- "Short-term": a few hundred days (partly quasi-periodic), \\
- "Anomalous Lows": about 5.4 years (quasi-periodic),  and \\
- "Long-term variability": 20-30\,yrs. \\

The cycle-to-cycle variability has been covered above, it is largely
governed by the described "rules" about the preferred binary phases for the
\emph{Turn-On} (Fig.~\ref{fig:bin_phase}), the length of the 35-day cycles (Fig.~\ref{fig:length}),
the (greatly) forbidden changes of the length from cycle to cycle, and the  corresponding
"jumps" in the (O - C)-diagram (Fig.\ref{fig:O-C_start}).

The "short-term" variability with "ups" and "downs" in the (O - C)-diagram, over tens
of 35-day cycles (a few hundred days) is already evident from Fig.~\ref{fig:O-C_start}.
We repeat the diagram (with additional informations) again in 
Fig.~\ref{fig:O-C_start2}, in order to emphasize that the O-C data can be 
piecewise described by linear functions, representing a constant
cycle length for tens of cycles, e.g. 21 x $P_{orb}$ ("up"), 20.5 x $P_{orb}$ ("flat")
and 20 x $P_{orb}$ ("down"). Apart from a long-term trend upward, a significant
quasi-periodic behavior is evident.

\subsection{Anomalous Lows \label{sec:AL}}
Medium-term variations (5-10 years) are observed in the form of \emph{Anomalous Lows}.
Soft X-rays (at least up to $\sim 7\,keV$) are absorbed, such that most monitoring instrument
cannot follow the evolution of flux or pulse period. It is believed, that the absorption is
produced by the accretion disk, either by blowing up their outer parts
or by substantially changing the tilt of the accretion disk, intercepting the X-ray beam.

\begin{figure}
\includegraphics[angle=0,width=10cm]{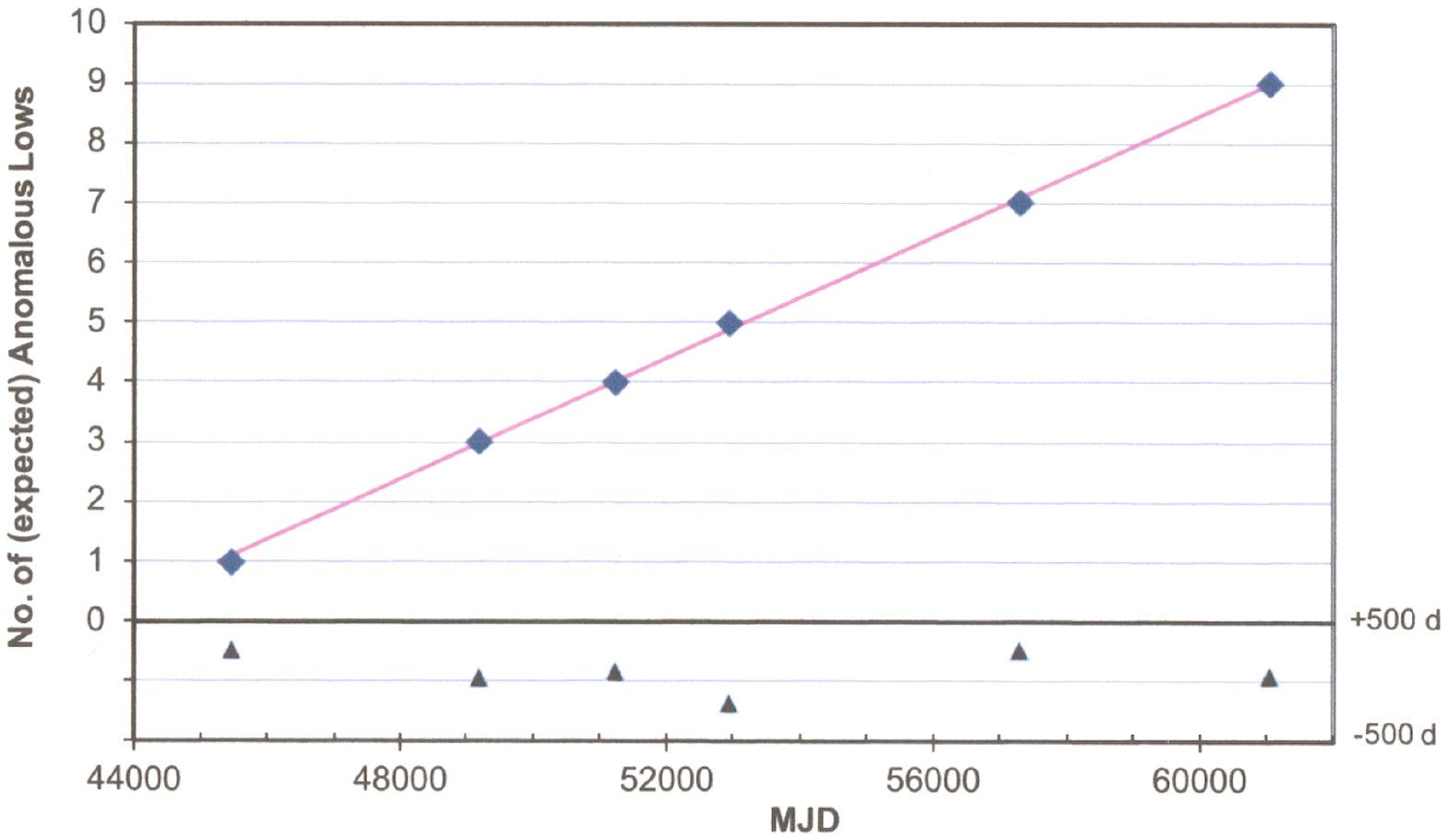}
\vspace{-8mm}
\caption{The counting numbers of  "expected" \emph{Anomalous Lows}  (ALs) versus
the start times  (in MJD) of actually observed ALs (see Table~\ref{tab:AL}).
A linear fit (allowing some expected
ALs (nos. 2, 6 and 8) not to have been realized) leads to a mean period of $(5.35 \pm0.03)\,yr$.
The residuals (observed - expected) are shown in the lower part (right hand ordinate, in days)
ranging from -215 \,d to +200\,d, a deviation from a strict periodicity of maximal 11\%.
}
\label{fig:ALs_plot}
\end{figure}

We refer to \citet{Staubert_etal17}, reproduce their Table~2 and add the most recent
event from early 2026 (Table~\ref{tab:AL}). The timing of the six events
observed so far is consistent with a repetition period of $\sim 5.4\,yr$, but with the
provision that some of the (expected) events did not actually materialize,
which seems to be the case three times out of nine. Fig.~\ref{fig:ALs_plot}
demonstrates that a mean "principle appearance period" of $(5.35~\pm~ 0.03)\,yr$ 
(taking the times of the onset of the ALs,  with a constant uncertainty of $\pm 35$\,d)
describes the data surprisingly well. The scatter in occurrence time with respect
to a strict periodicity is of the order of maximal 12\%.

We like to mention a few general points connected with the
appearance of an \emph{Anomalous Low} (see also \citet{Staubert_etal17}): 
\begin{itemize}
\item X-ray light curves show that the flux is gradually decreasing before such
an event, actually for several 35-day cycles. This is very prominently observed 
for AL3 \citep{Coburn_etal00} and for the currently on-going AL6 (our 
analysis of data from monitoring instruments). 
\item Strong spin-down (a rapid increases of the rotational period) of the neutron star 
is observed for AL3 \citep{Staubert_etal09a} and AL6 (our analysis of recent 
\textsl{NuSTAR} observations). 
\item The duration of the 35-day cycle tends to be shorter than average, clearly
seen in AL3 \citep{Staubert_etal09a}, and AL6 (again based on our analysis of
the first \textsl{NuSTAR} observations of the current AL).
This seems to be in line with theoretical expectations: when the tilt of the disk 
becomes smaller, the precessional period should decrease
(private communication  Susanne Schandl). 
\item In Fig.~\ref{fig:O-C_all} a strong drop in (O - C) is seen around 
MJD 52000, apparently associated with AL3 (and in correlation with a dramatic 
spin-down of the neutron star). This is the consequence of the observed short
duration of 35-day cycles during this AL and the cycle counting bridging this AL
adopted by \citet{Staubert_etal09a}. Here it is argued that the accretion disk
actually did one cycle more than expected with the usual length of around 
20.5 x P(orb). Physically this means that the assumed synchronization between
the precessional periods of the neutron star and of the accretion disk did break
down.  \citet{Staubert_etal09a} has called it a sign that the accretion disk 
has a ``life of its own, and is able to temporarily escape the "slaving" by the 
neutron star. After the end of AL3 the coupling was found to be re-established.
However, \citet{Staubert_etal13} showed that it is possible to follow the neutron star
precession by "pulse profile fitting", making use of the systematic changes of the 
pulse profiles, allowing to determine \emph{Turn-On} times. This led to the conclusion
that the two precessions - neutron star and accretion disk - are generally very well
synchronized, both showing the same irregular behavior (see also \citet{Postnov_etal13}).
If we follow the notion of the neutron star being the "master clock" and the 
synchronization does not break down during an AL,  the observed evolution 
of the pulse  profiles require that also the neutron stars precessional period is
much shorter during ALs (see the note above on the "pulse profile fitting" 
for the observations during the very recent AL6). A short neutron star free 
precessional period is possible in the case of three-axial precession \citep{Shakura_etal98}.
\item The long-term variability, above associated with a time scale of 20-30\,yrs
could well be linked to the quasi-period of the lows: 4 x 5.4\,yr = 21.6\,yr. 
\item Looking at the two prominent long-term sections in (O-C) (Figs.~4, 13),
the currently on-going low (AL6), which started in February 2026, has the potential 
of lasting very long. AL3 lasted for 602\,days! 
\end{itemize}

\begin{table}
\caption[]{Information about all observed  (col.~1) \textsl{Anomalous Lows} in Her~X-1.
\label{tab:AL}
The \textsl{Turn-On} times given are for the last observed \textsl{Turn-On} before the AL,
and the first \textsl{Turn-On} after the end of the AL.
Note that with the numbering of col.~2, the mean repetition period of the ALs is 
$\sim5.35$\,yrs (assuming that nos. 2, 6 and 8 were not observed). AL6 is currently ongoing.
The 35\,d cycle numbering is according to \citet{Staubert_etal83}.
This table is an update of Table~2 of \citet{Staubert_etal17}.}
\vspace{-0.8cm}
\begin{center}
\begin{tabular}{llllllll}
\hline\noalign{\smallskip}
Obse  & No.   &  No.    & Turn-     & No.     & Turn-    & Dur    & Ref.       \\
rved   & of     & 35\,d   & On          & 35\,d  & On        & ation   &              \\
ALs   & exp    & cycl.    &                & cycl.   &             &  of      &              \\
         & ect.    & before &                & after   &             &  AL     &              \\
         & ALs    & AL       & [MJD]      & AL      & [MJD]  & [d]      &              \\
\hline\noalign{\smallskip}
AL 1   & 1       & 119     & 45469     & 127    & 45752    & 283   & 1,2       \\
AL 2   & 3       & 226     & 49205     & 229    & 49308    & 103   & 3          \\
AL 3   & 4       & 284     & 51224     & 302    & 51826    & 602   & 4,5,6,7 \\
AL 4   & 5       & 334     & 52945     & 340    & 53157    & 212   & 8          \\
AL 5   & 7       & 459     & 57302     & 461    & 57372    & 70     & 8          \\
AL 6   & 9       & 567     & 61045     & on-     & going     & ?       & 9          \\	
\noalign{\smallskip}\hline
\end{tabular}\end{center}
\vspace{-0.2cm}
References: 1. \textsl{EXOSAT}: \citet{Parmar_etal85}, 2. optical: \citet{Delgado_etal83}, 
3. \textsl{Rosat / ASCA / BATSE / EUVE}:  \citet{Vrtilek_etal94}, 4. \textsl{RXTE}/ASM: 
\citet{LevineCorbet_99}, 5. \textsl{Beppo}SAX: \citet{Parmar_etal99}, 6. \textsl{RXTE}: 
\citet{Coburn_etal00}, 7. \textsl{RXTE}: \citet{Still_etal01a}, 
8. \textsl{RXTE}/ASM / \textsl{Swift}/BAT: \citet{Staubert_etal17},
9. R. Staubert, this work.
\label{tab:AL}
\end{table}

\vspace{-3mm}
\subsection{Long-term variations }
\begin{figure}
\includegraphics[width=10.5cm]{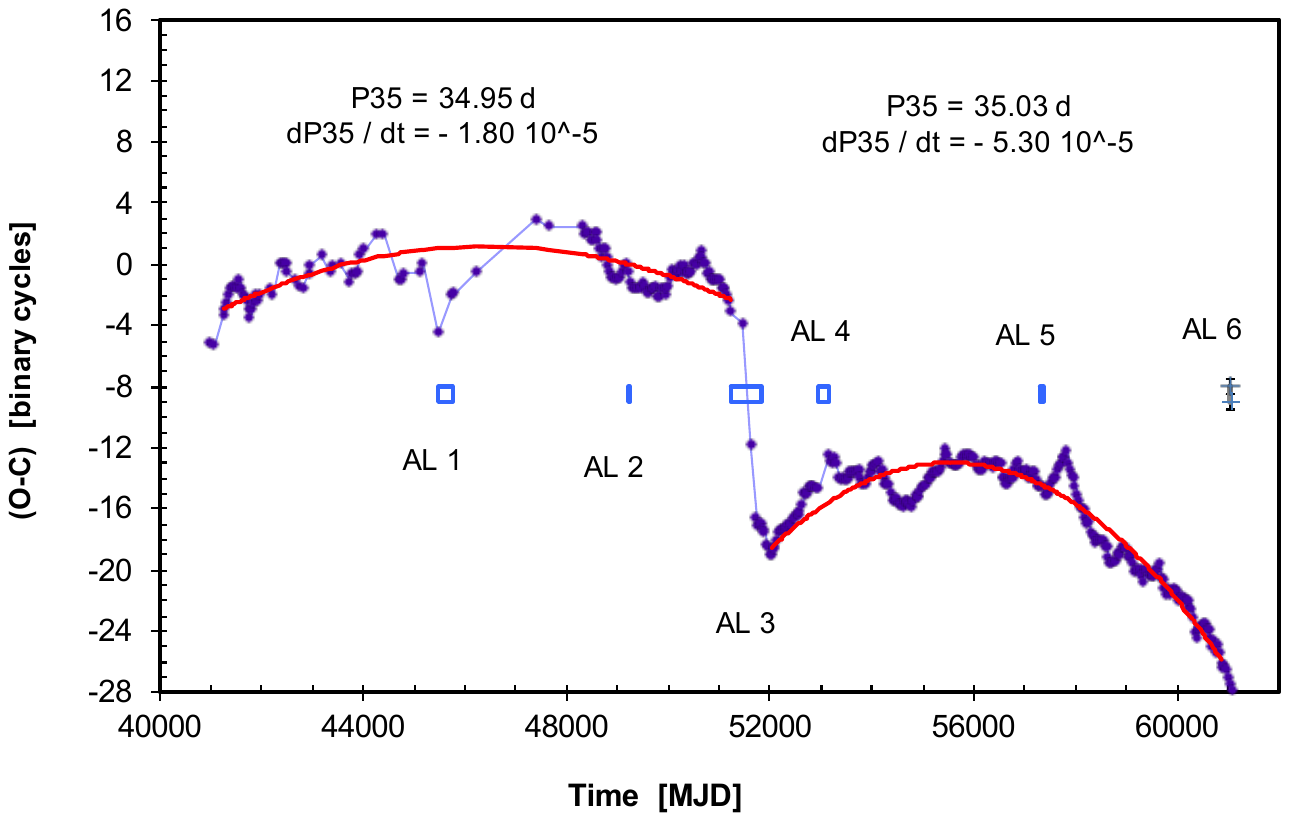}
\vspace{-13mm}
\caption{O-C diagram (as in Fig.~\ref{fig:O-C_all}) showing that there are two distinct
time periods of around 30 yrs, separated by the 602 day long "Anomalous 
Low" (AL3) centered on 23 March 2000 (MJD 51626).  The course evolution
can be described by parabolic functions providing a measure of the
long-term change of the value of the 35-day period of the order of a 
few times 10$^{-5}$ d/d. In addition there is a fine structure demonstrating
stronger variations on various shorter time scales.
}
\label{fig:parabolics}
\label{fig:longterm_all}
\end{figure}

In Fig.~\ref{fig:O-C_all} two groups of data, separated by a
dramatic jump around MJD 52000 in (O-C), near the the 602\,d long 
\emph{Anomolous Low} (AL3), are evident. We repeat this figure in 
Fig.~\ref{fig:longterm_all} with a crude attempt  to model the 
long-term evolution in both sections with parabolic functions
(shown in red). Both functions start off with a long length of the
35-day modulation (near 21 x $P_{orb}$), and follow a parabolic
paths.
For the section before AL3 the best-fitting values are P = 34.95\,d and 
dP/dt = ($-1.8 \pm 0.2$) x $10^{-5}\,\rm s\,s^{-1}$,
and for the section after AL3 the values are P = 35.03\,d and 
dP/dt = ($-5.30 \pm 0.23$) x $10^{-5}\,\rm s\,s^{-1}$.
Both sections end at an \emph{Anomalous Low} after the 
X-ray flux showed a gradual decrease (for AL3 see e.g. 
\citet{Coburn_etal00}). Here we do not want to discuss this 
interesting phenomenon further since it is the main subject
of a parallel paper describing the analysis of the 35-day 
X-ray flux modulation using full light curve data from all the 
monitoring instruments \citep{Revnivtsev_etal26}. In this paper 
a physical mechanism for a parabolic evolution of (O-C) is proposed.

\begin{figure}
\includegraphics[width=10.5cm]{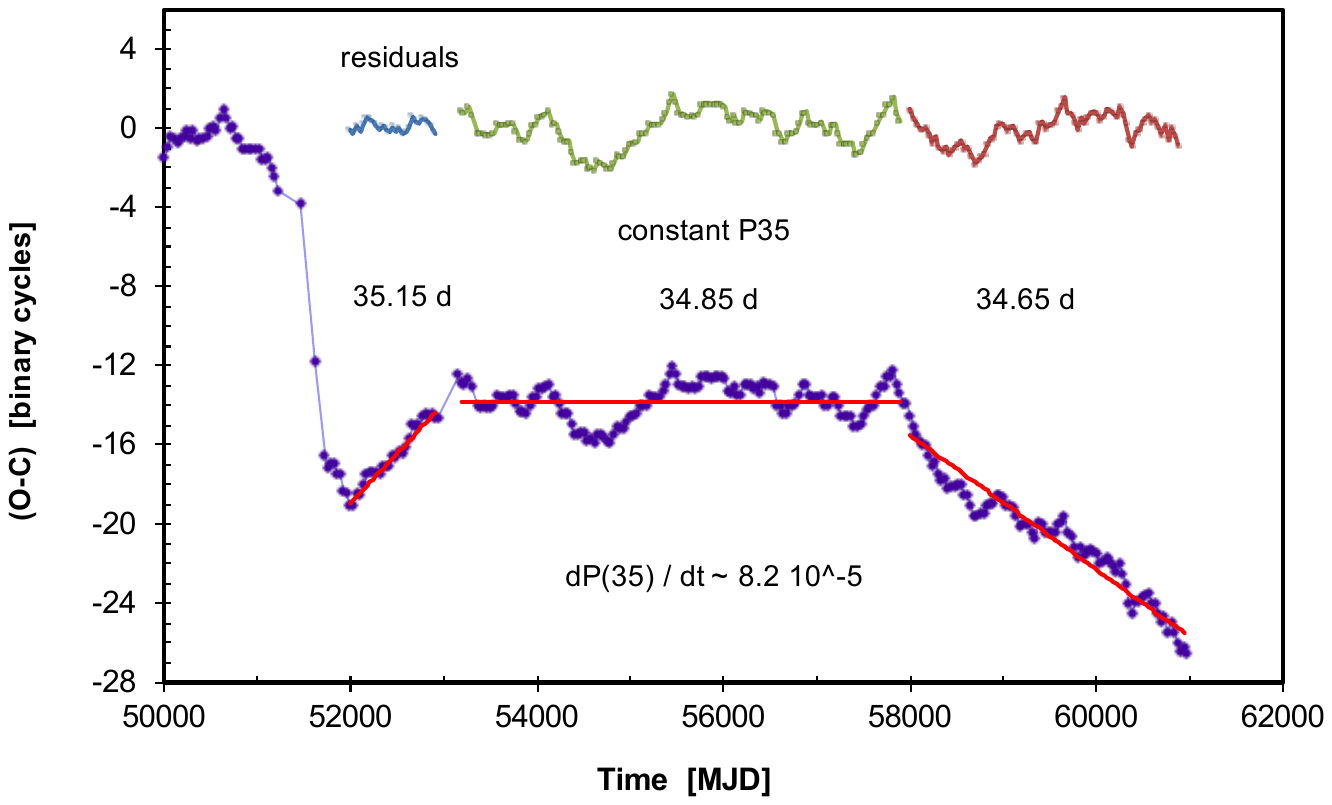}
\vspace{-13mm}
\caption{O-C diagram for the time after AL3. Here the long-term 
evolution of the precessional period is modeled by three sections
with mean constant periods (20.67, 20.50 and 20.34 binary
cycles. The short-term variability is evident.
}
\label{fig:longterm_second}
\end{figure}

For the section after AL3 (> MJD 52000) we offer here an alternative 
description (see Fig.~\ref{fig:longterm_second}): three intervals with 
constant mean values (note that the effective dP(35)/dt, from the first 
to the third interval, is -8.2 x $10^{-5}\,\rm s\,s^{-1}$, an even larger 
value as for the parabolic solution).
From a physical point of view this solution may be attractive, since 
the times of the break points between the intervals could  be
associated with abrupt changes in the system, possibly
detectable in other observables. Apart from the known correlations
between pulse period and X-ray flux and between pulse period
and (O-C), there are likely to be other correlations, possibly even with 
the values of the cyclotron line.
Of course, there is strong short term variability, in part quasi-periodic,
the investigations of which offer new insight.
We just note here that it is conceivable that also the very long-term
behavior might be of quasi-periodical nature - with a  period
around 20 to 30 yrs. \\

\vspace{-3mm}
\section{Discussion \label{sec:Discussion}}
With regard to the here presented collection of X-ray \emph{Turn-Ons}
(based on observations over 55\,yrs), the main questions are the following:
\begin{itemize}
\item What causes the accretion disk to tilt (with respect to the binary orbit),
to warp and to precess? 
\item How is the physical shape and movement of the disk best described? 
\item What is the physical origin of the irregularity of the 35-day modulation? 
\item Does the neutron star itself show free precession and is it
synchronized to that of the accretion disk? 
\item What exactly happens during Anomalous Lows? 
\item Is the long-term evolution of the accretion disk precession, as
represented by the (O - C) diagram, the consequence of derivative of
the neutron star precession period due to continued accretion
\citep{Revnivtsev_etal26}?
\end{itemize}

There is a vast amount of literature in which these questions are addressed
in detail. Here we will give a short summary (relying on many references), 
and referring only to the most fundamental aspects.

Let us start of with some very general statements about the Her X-1 / HZ-Her system: \\
The eclipsing binary X-ray pulsar system Her X-1 / HZ-Her is one of the best 
studied system of its kind. For a detailed description see,  e.g. \citet{Staubert_etal09}.
The relevant physical features for this work are a strong X-ray flux from the
highly magnetized neutron star Her~X-1, powered by accretion of material from 
its optical companion HZ~Her trough a tilted and warped accretion disk.
It is believed that the 35-day modulation of the X-ray flux is caused by
the variable obscuration of the X-ray source (small regions near the magnetic poles
of the neutron star) by the precessing accretion disk. This physical picture was first 
proposed by \citet{Katz_73}  and further discussed by \citet{Roberts_74}.
Today there is a vast amount of literature dealing with describing and physical modeling 
of the system.

Through modeling one can understand the overall 35-day X-ray light curve and why
the \emph{Turn-Ons} happen at the preferred orbital phases 0.25 and 0.75
\citep{CrosaBoynton_80,BoyntonCrosaDeeter_80,Shakura_etal99,StillBoyd_04,Klochkov_etal06,
LevineJernigan_82}.
This includes the binary eclipses and the (so far not discussed) pre-clipse
dips (with an effective periodicity of 1.62\,d) through obscuration by the accretion 
stream which transports the material from the optical star to the accretion disk. 
Physically, this stream plays an important role for the properties of the
accretion disk (there is a beat frequency correlation between the orbit and the 
35-day precession).

A crucial point is the assumption of free precession of the neutron star, originally
proposed by \citet{Ruderman_70} as explanation of the 35-day modulation.
It has then be shown, that in fact the accretion disk is responsible for this
modulation. But we now believe  that the neutron star in Her~X-1 is indeed
precessing, in fact with the same period as the accretion disk, namely close to 
35\,days. This is quite remarkable, and cannot be accidental.
Our current understanding is that there is most likely a strong physical coupling
between the neutron star and the accretion disk through magnetic interaction
at the inner edge of the accretion disk and through the X-ray beam heating the
atmosphere of HZ~Her, modulating  the accretion stream, which then acts on
the outer parts of the accretion disk by tidal forces and dynamical actions.
The neutron star precession provides a long-term clock mechanism 
\citep{Shakura_etal99}, see also the discussion in \citet{Staubert_etal09a,Revnivtsev_etal26}.
This scenario has received increasing support through observational evidence
for the reality of neutron star free precession, mainly by studying the 
evolution of the pulse period and the systematic variations of the X-ray
pulse profiles with 35-day phase 
\citep{Truemper_etal86,Soong_etal87,Soong_etal90,Petterson_etal91,Deeter_etal98,
ScottLeahyWilson_00,Staubert_etal09a,Staubert_etal13,Postnov_etal13,Shakura_etal18,Kolesnikov_etal22}. 
See in particular \citet{Staubert_etal13}: here times of \emph{turn-ons} found by 
two different methods - fitting the X-ray light curve and using the variable
shape of the pulse profiles (as described above) - are compared, with the conclusion
that \textsl{the "pulse profile clock" is just as irregular as the "turn-on clock".
Both clocks appear to be perfectly synchronized}. The conclusion is:
there is only \textsl{one 35-day clock} in the system. \\

So, the final questions are then what is the physical reason for the demonstrated
\emph{irregularity of the 35-day clock} and its \emph{long-term evolution}? \\

Regarding the irregularity, we are not able to pin-point to a particular physical 
reason, rather give a very general argument about the inherent and un-avoidable
variability in the system. Her X-1 / HZ Her is a complex
\emph{"cosmic machine"} with many feedbacks (actions and re-actions) 
between all components of the system, as clearly seen in correlations
between various observables  \citep{Staubert_etal00,Klochkov_etal09}.
The optical star has a life and variability of its own, both intrinsically and
in coupling with the neutron star. Some
part of the X-ray beam created on the neutron star hits the atmosphere of 
HZ Her and heats it, leading to increased mass loss, hence variable mass
transfer rates through the accretion stream, which hits the „fluffy“ (non-rigid)
accretion disk (about midway between its inner and outer edge), leading to a 
warped, inclined and precessing accretion disk. This disk and its hot corona 
sending off a strong coronal wind \citep{SchandlMeyer_94} then 
shades - precession phase dependent - the X-ray emitting spots and accretion 
columns at the magnetic poles of the neutron star. The precessing neutron star
produces variable pulse profiles.  A variable accretion rate onto the magnetic 
poles (facilitated through the interaction of the magnetosphere of the neutron 
star with the inner edge of the accretion disk) leads to a variable flux and a 
variable transfer of angular momentum, observed as spin-up/spin-down 
(variations in pulse period) of the pulsar.  In summary, it appears not surprising
that the combination of all these physical actions and re-actions lead a
variable observational appearance. \\

\vspace{-4mm}
\section{Summary \label{sec:Summary}}
We present the results of a study of 55-years of observations of Her X-1:
Table~\ref{bigtable} gives a complete set of times (in MJD) of the  X-ray 
\emph{Turn-Ons} (until cycle number 570, April 2026).
The following main features are summarized here:
\begin{enumerate}
    \item The \emph{Turn-Ons} clearly show preference to occur at orbital phases 
    $\sim 0.25$ and $\sim 0.75$ (see Fig.~\ref{fig:bin_phase}). This can be understood
    by the model of tidal wobbling of a tilted precessing accretion disk.
    \item The duration of the 35-day cycle clusters around three discrete values:
    20.0, 20.5 and 21.0 orbital cycles with a dominant peak at  $20.5 \times P_{\text{orb}}$ 
   ($\sim 57\%$) and smaller peaks for the two other values (Fig.~\ref{fig:length}).
   This discretization (already evident in the first X-ray measurements of Her X-1 (Fig. \ref{fig:O-C_start}))
   conforms with the tidal wobbling of the precessing tilted accretion disk.
    \item The long-term trend in the 35-day period duration is seen in the (O-C) diagram 
    (Fig.~\ref{fig:O-C_all} and Fig. \ref{fig:longterm_all}). There are two sections, separated by the 
    long \emph{Anomalous Low}
    of 1999/200 (AL3), which can be modeled by parabolic functions with a smooth evolution of 
    the mean 35-day period (Fig. \ref{fig:longterm_all})) (\citet{Revnivtsev_etal26}, offers
    a physical interpretation).
  \item There are short-term (hundreds of days) variations (in part quasi-periodic) overlaid to 
    the long-term trend. And straight "ups" and "downs" in the O-C diagram demonstrate rather rapid
    ($\sim 35\,d$) period changes (Fig.~\ref{fig:O-C_start2}). 
    \item A medium-term variation is observed in the form of \emph{Anomalous Lows}, an 
    essential shut-off (at least of soft) X-rays, with an apparent underlying period of ($\sim 5.4\,yr$),
    but with the ability to not realize some of the lows (Table~\ref{tab:AL} and Fig.~\ref{fig:ALs_plot}).
\end{enumerate}

\subsection{Acknowledgements} 
 We are thankful for the possibility to use the NASA archive HEASARC
 at Goddard Spaceflight Center in retrieving archival data of those
 missions that monitored the flux of Her~X-1. The same holds for the
 archive of Maxi data, available at the Japanese institute RIKEN.
 We thank all the students and coworkers at the University of T\"ubingen
 who have - over many years - helped to fill and maintain our table of the
 times of \emph{Turn-On} of Her~X-1, in particular Mathias Kunz, J\"orn Wilms
 and Dmitry Klochkov. We also acknowledge the help by the informatics 
 student Tim Staubert (grand son of the first author) in preparing various 
 routines to analyze data taken from the relevant archives. We extend our
 thanks to those colleagues who have made use of the data provided by us
 and referenced it under "private communication R\"udiger Staubert".
 Vladislav Revnivtsev was partly supported by RSF grant 25-22-00295.
 The authors thank Prof. N.I. Shakura for useful discussions.

\bibliographystyle{aa}
\vspace{-3mm}
\bibliography{refs_herx1_May26}

\newpage

\appendix
\section{Table of Turn-Ons}

\begin{table*}
\vspace*{2mm}
\caption{Table of times of the 35-day \emph{Turn-Ons} of Her X-1 (on-line or on-demand available: staubert@astro.uni-tuebingen.de)}
\label{bigtable}
\centering
\begin{tabular}{l l l l l l l l l}
\hline
35\,d         &  Turn-On  &  +/-       & orbital         & orbital        & mean$^{4}$  & (O - C)$^{2,}$$^{3}$   & Satellite /     & Reference / \\
cycle         &                 &             & cycle      & phase$^{2}$   & length          &                                      & Instrument   & Comment  \\
no.$^{1}$  &  [MJD]      &  [d]       & no.              &                   & [bin cyc]      & [bin cyc]                         &                     &                  \\
  \hline	  															
-15	&	40797,3	&	0,8	&	-3272	&	0,22	&	20,6	&	-5,5	&	opt.	        &	29  \\ 
-10	&	40972,2	&	0,5	&	-3169	&	0,09	&	20.5	&	-5,1	&	opt.		&	29 \\ 
-7	&	41076,6	&	0,3	&	-3108	&	0,50	&	20,9	&	-5,2	&	opt.	        &	29 \\ 
-2	&	41254,1	&	0,4	&	-3004	&	0,90	&	20,8	&	-3,3	&	UHURU    &	3  \\
-1	&	41289,5	&	0,02	&	-2983	&	0,73	&	21,0	&	-3,0	&	UHURU	&	1, 2 \\
0	&	41325,2	&	0,06	&	-2962	&	0,71	&	21,0	&	-2,5	&	UHURU	&	1, 3, 7+ \\
1	&	41360,9	&	0,04	&	-2941	&	0,70	&	21,0	&	-2,0	&	UHURU	&	1 \\
2	&	41396,5	&	0,04	&	-2920	&	0,66	&	20.6	&	-1,6	&	UHURU	&	1, 3 \\
3	&	41431,6	&	0,07	&	-2899	&	0,30	&	20,5	&	-1,4	&	UHURU	&	1 \\
4	&	41466,4	&	0,05	&	-2879	&	0,75	&	20,5	&	-1,5	&	UHURU	&	1 \\
5	&	41501,2	&	0,04	&	-2858	&	0,21	&	20,5	&	-1,5	&	UHURU	&	1, 3 \\
6	&	41536,0	&	0,04	&	-2838	&	0,70	&	21,0 &	-1,5	&	UHURU	&	1 \\
7	&	41571,6	&	0,02	&	-2817	&	0,65	&	20.0 &	-1,1	&	UHURU	&	1 \\
8	&	41605,6	&	0,04	&	-2797	&	0,64	&	20,1	&	-1,6	&	UHURU	&	1 \\
9	&	41639,7	&	0,03	&	-2777	&	0,71	&	20,5	&	-2,0	&	OSO 7	&	1, 4, 5 \\
10	&	41674,5	&	0,03	&	-2756	&	0,17	&	20,1	&	-2,1	&	UHURU	&	1, 3, 4\\
11	&	41708,7	&	0,07	&	-2736	&	0,28	&	20,0	&	-2,5	&	UHURU	&	1 \\
12	&	41742,7	&	0,04	&	-2716	&	0,30	&	19.9	&	-2,9	&	UHURU	&	1 \\
13	&	41776,6	&	0,05	&	-2696	&	0,24	&	21.0	&	-3,5	&	 opt.	&	6    \\
14	&	41812,3	&	0,05	&	-2675	&	0,23	&	21,0	&	-3,0	&	 opt.	&	6    \\
15	&	41848,0	&	0,05	&	-2654	&	0,23	&	21,0	&	-2,5	&	 opt.	&	6    \\
16	&	41883,7	&	0,05	&	-2633	&	0,24	&	20,0	&	-2,0	&	 opt.	&	6    \\
17	&	41917,8	&	0,41	&	-2613	&	0,27	&	21,0	&	-2,5	&	UHURU	&	1 \\
18	&	41953,4	&	0,05	&	-2592	&	0,24	&	20,6 &	-2,0	&	 opt.	&	6    \\
24	&	42163,2	&	0,02	&	-2469	&	0,64	&	20,3	&	-1,6	&	Copernicus	&	1, 7+\\
26	&	42232,3	&	0,30	&	-2428	&	0,26	&	21.0	&	-2,0	&	UHURU	&	29 \\ 
30	&	42375,1	&	0,10	&	-2344	&	0,25	&	20.5	&	0,0	&	UHURU	&	1 \\
31	&	42409,9	&	0,10	&	-2324	&	0,74	&	20,5	&	0,0	&	UHURU	&	1 \\
32	&	42444,8	&	1,00	&	-2303	&	0,24	&	20,5	&	0,0	&	Ariel 5	&	8*  \\
33	&	42479,6	&	0,07	&	-2283	&	0,74	&	20,0	&	0,0	&	Ariel 5	&	1, 7+ \\
34	&	42513,6	&	0,02	&	-2263	&	0,71	&	20,4	&	-0,5	&	Copernicus	&	1, 7+ \\
38	&	42652,2	&	0,04	&	-2181	&	0,23	&	20,3	&	-1,0	&	 OSO 8	&	1, 9, 4 \\
41	&	42755,9	&	0,10	&	-2120	&	0,23	&	20,5	&	-1,5	&	UHURU	&	2 \\
43	&	42825,5	&	0,10	&	-2079	&	0,16	&	20,8	&	-1,6	&	Ariel 5	&	1 \\
46	&	42931,7	&	0,50	&	-2017	&	0,63	&	21.1	&	-0,6	&	Ariel 5	&	1 \\
47	&	42967,5	&	0,05	&	-1996	&	0,69	&	20.6	&	-0,1	&	SAS 3	&	1 \\
53	&	43177,6	&	0,04	&	-1872	&	0,28	&	20,3	&	0,5	&	Ariel 5	&	1, 10\\
58	&	43350,1	&	0,05	&	-1771	&	0,71	&	20,9	&	-0,5	&	 opt.	&	6   \\
59	&	43385,7	&	0,1	&	-1750	&	0,66	&	20,5	&	-0,1	&	 OSO 8	&	1, 9, 11\\
64	&	43560,1	&	0,1	&	-1647	&	0,24	&	20,3	&	0,0	&	HEAO-1 A4	&	12, 13* \\
69	&	43732,5	&	0,3	&	-1546	&	0,61	&	21.0	&	-1,1	&	HEAO-1 A4	&	1, 13*, 11 \\
70	&	43768,2	&	0,5	&	-1525	&	0,64	&	20.5	&	-0,6	&	HEAO-1 A4	&	13, 14*  \\
72	&	43837,9	&	0,2	&	-1484	&	0,63	&	20,6	&	-0,6	&	Ariel 5	&	1 \\
73	&	43873,0	&	0,2	&	-1463	&	0,28	&	21.0	&	-0,5	&	Ariel 5	&	1 \\
75	&	43944,5	&	0,5	&	-1421	&	0,33	&	20.7	&	0,6	&	Einstein MPC	&	15 \\
77	&	44014,9	&	0,9	&	-1380	&	0,71	&	20,6	&	1,0	&	 opt.	&	6   \\
84	&	44260,5	&	0,9	&	-1235	&	0,20	&	20,5	&	2,0	&	Ariel 5	&	1 \\
88	&	44400,0	&	0,9	&	-1153	&	0,23	&	20,2	&	2,0	&	 opt.	&	6   \\
97	&	44708,5	&	0,9	&	-972	        &	0,71	&	20,5	&	-1,0	&	 opt.	&	6   \\
98	&	44743,4	&	0,9	&	-951	        &	0,23	&	20,9	&	-1,0	&	 opt.	&	6   \\
99	&	44779,0	&	1	&	-930	        &	0,17	&	20,5	&	-0,6	&	Ariel 6	&	1 \\
109	&	45127,6	&	0,9	&	-725.   	&	0,23	&	21.0	&	-0,5	&	 opt.	&	6   \\
110	&	45163,3	&	0,9	&	-704   	&	0,23	&	20,0	&	0,0	&	 opt.	&	6   \\
119	&	45469,4	&	0,9	&	-524		&	0,23	&	20.8	&	-4,5	&	 opt.	&	17, AL1 \\
127	&	45752,5	&	1	&	-358		&	0,76	&	20,6	&	-2,0	&	EXOSAT	&  19, AL1 \\
128	&	45787,5	&	0,5	&	-337		&	0,34	&	20,6	&	-1,9	&	EXOSAT	&  19 \\
140	&	46208,0	&	0,5	&	-90		&	0,67	&	20,6	&	-0,6	&	EXOSAT	&  19, 20 \\
\hline
\end{tabular} 
\vspace*{1mm} \\
\end{table*}

\begin{table*}
\vspace*{2mm}
\addtocounter{table}{-1}
\caption{continued (2)}
\label{bigtable_ext2}
\centering
\begin{tabular}{l l l l l l l l l}
\hline
35\,d       &  Turn-On    &  +/-     & orbital            & orbital      & length of           & (O - C)     &instrument  & Reference / \\
cycle no. &  [MJD]        &  [d]     & cycle no.       & phase       & cycle [bin cyc]  & [bin cyc]   &                  &  Comment \\
  \hline	  															
174	&	47399,0	&	0,3	&	611	&	0,19	&	20,43	&	2,95	&	Ginga	&	21* \\
181	&	47642,2	&	0,2	&	754	&	0,24	&	20,50	&	2,49	&	Ginga	&	22* \\
200	&	48304,4	&	0,05	&	1143	&	0,70	&	20,22	&	2,46	&	 ROSAT	&	23* \\
202	&	48373,1	&	0,4	&	1184	&	0,14	&	20,65	&	1,89	&	BATSE	&	24 \\
203	&	48408,2	&	0,4	&	1204	&	0,78	&	20,53	&	2,04	&	BATSE	&	24 \\
204	&	48443,1	&	0,4	&	1225	&	0,31	&	19,94	&	2,07	&	BATSE	&	24 \\
205	&	48477,0	&	0,4	&	1245	&	0,25	&	20,53	&	1,51	&	BATSE	&	24 \\
206	&	48511,9	&	0,4	&	1265	&	0,77	&	21,06	&	1,53	&	BATSE	&	24 \\
207	&	48547,7	&	0,4	&	1286	&	0,83	&	20,47	&	2,09	&	BATSE	&	24 \\
208	&	48582,5	&	0,4	&	1307	&	0,30	&	20,00	&	2,06	&	BATSE	&	24 \\
209	&	48616,5	&	0,4	&	1327	&	0,30	&	20,06	&	1,56	&	BATSE	&	24 \\
210	&	48650,6	&	0,4	&	1347	&	0,35	&	20,06	&	1,05	&	BATSE	&	24 \\
211	&	48684,7	&	0,4	&	1367	&	0,41	&	20,59	&	0,49	&	BATSE	&	24 \\
212	&	48719,7	&	0,5	&	1388	&	0,00	&	20,76	&	0,93	&	BATSE	&	24 \\
213	&	48755,0	&	0,5	&	1408	&	0,76	&	20,00	&	1,02	&	BATSE	&	24 \\
214	&	48789,0	&	0,4	&	1428	&	0,76	&	19,88	&	0,52	&	BATSE	&	24 \\
215	&	48822,8	&	0,5	&	1448	&	0,64	&	20,12	&	-0,10	&	BATSE	&	24 \\
216	&	48857,0	&	0,5	&	1468	&	0,75	&	20,06	&	-0,49	&	BATSE	&	24 \\
217	&	48891,1	&	0,4	&	1488	&	0,81	&	20,47	&	-0,93	&	BATSE	&	24 \\
218	&	48925,9	&	0,5	&	1509	&	0,28	&	20,41	&	-0,96	&	BATSE	&	24 \\
219	&	48960,6	&	0,4	&	1529	&	0,69	&	20,59	&	-1,05	&	BATSE	&	24 \\
221	&	49030,6	&	0,4	&	1570	&	0,86	&	20,53	&	-0,88	&	BATSE	&	24 \\
222	&	49065,5	&	0,4	&	1591	&	0,39	&	20,88	&	-0,85	&	BATSE	&	24 \\
223	&	49101,0	&	0,4	&	1612	&	0,27	&	20,88	&	-0,47	&	BATSE	&	24 \\
224	&	49136,5	&	0,4	&	1633	&	0,15	&	20,53	&	-0,09	&	BATSE	&	24 \\
225	&	49171,4	&	0,4	&	1653	&	0,68	&	20,06	&	0,05	&	BATSE	&	24 \\
226	&	49205,5	&	0,4	&	1673	&	0,73	&	19,88	&	-0,51	&	BATSE	&	24 \\
227	&	49239,3	&	0,4	&	1693	&	0,61	&	20,26	&	-1,13	&	BATSE, AL2	&	24 \\
229	&	49308,2	&	0,4	&	1734	&	0,14	&	20,47	&	-1,60	&	BATSE	&	24 \\
230	&	49343,0	&	0,5	&	1754	&	0,61	&	20,53	&	-1,63	&	BATSE	&	24 \\
231	&	49377,9	&	0,4	&	1775	&	0,14	&	20,47	&	-1,61	&	BATSE	&	24 \\
232	&	49412,7	&	0,4	&	1795	&	0,60	&	20,65	&	-1,64	&	BATSE	&	24 \\
233	&	49447,8	&	0,4	&	1816	&	0,25	&	20,35	&	-1,49	&	BATSE	&	24 \\
234	&	49482,4	&	0,4	&	1836	&	0,60	&	20,82	&	-1,64	&	BATSE	&	24 \\
235	&	49517,8	&	0,4	&	1857	&	0,42	&	20,35	&	-1,14	&	BATSE	&	24 \\
236	&	49552,4	&	0,4	&	1877	&	0,77	&	20,06	&	-1,47	&	BATSE	&	24 \\
237	&	49586,5	&	0,4	&	1897	&	0,83	&	20,53	&	-1,91	&	BATSE	&	24 \\
238	&	49621,4	&	0,8	&	1918	&	0,36	&	20,76	&	-1,88	&	BATSE	&	24 \\
239	&	49656,7	&	0,5	&	1939	&	0,12	&	20,53	&	-1,62	&	BATSE	&	24 \\
240	&	49691,6	&	0,4	&	1959	&	0,65	&	20,47	&	-1,59	&	BATSE	&	24 \\
241	&	49726,4	&	1,3	&	1980	&	0,12	&	20,35	&	-1,63	&	BATSE	&	24 \\
242	&	49761,0	&	0,4	&	2000	&	0,47	&	20,29	&	-1,48	&	BATSE	&	24 \\
243	&	49795,5	&	0,4	&	2020	&	0,76	&	20,41	&	-2,10	&	BATSE	&	24 \\
244	&	49830,2	&	0,4	&	2041	&	0,17	&	20,94	&	-2,07	&	BATSE	&	24 \\
245	&	49865,8	&	0,4	&	2062	&	0,11	&	20,35	&	-1,63	&	BATSE	&	24 \\
246	&	49900,4	&	0,4	&	2082	&	0,46	&	20,29	&	-1,61	&	BATSE	&	24 \\
247	&	49934,9	&	0,4	&	2102	&	0,75	&	20,88	&	-1,99	&	BATSE	&	24 \\
248	&	49970,4	&	0,4	&	2123	&	0,63	&	20,65	&	-1,61	&	BATSE	&	24 \\
249	&	50005,5	&	0,4	&	2144	&	0,28	&	21,00	&	-1,47	&	BATSE	&	24 \\
250	&	50041,2	&	0,5	&	2165	&	0,27	&	20,76	&	-0,97	&	BATSE	&	24 \\
251	&	50076,5	&	0,4	&	2186	&	0,04	&	20,65	&	-0,41	&	BATSE	&	24 \\
252	&	50111,6	&	0,4	&	2206	&	0,68	&	20,35	&	-0,56	&	BATSE	&	24 \\
253	&	50146,20	&	0,3	&	2227	&	0,03	&	20,65	&	-0,71	&	 ASM	&	25 \\
254	&	50181,30	&	0,2	&	2247	&	0,68	&	20,65	&	-0,56	&	 ASM	&	25 \\
255	&	50216,4	&	0,2	&	2268	&	0,32	&	20,59	&	-0,42	&	 ASM	&	25 \\
256	&	50251,40	&	0,4	&	2289	&	0,91	&	20,12	&	-0,10	&	 ASM	&	25 \\
257	&	50285,6	&	0,5	&	2309	&	0,02	&	21,12	&	-0,54	&	 ASM	&	25 \\
258	&	50321,5	&	0,3	&	2330	&	0,14	&	20,06	&	-0,10	&	 ASM	&	25 \\
259	&	50355,6	&	0,4	&	2350	&	0,20	&	20,53	&	-0,60	&	 ASM	&	25 \\
\hline
\end{tabular} 
\vspace*{1mm} \\
\end{table*}

\begin{table*}
\vspace*{2mm}
\addtocounter{table}{-1}
\caption{continued (3)}
\label{bigtable_ext3}
\centering
\begin{tabular}{l l l l l l l l l}
\hline
35\,d       &  Turn-On    &  +/-     & orbital            & orbital      & length of           & (O - C)     &instrument  & Reference / \\
cycle no. &  [MJD]        &  [d]     & cycle no.       & phase       & cycle [bin cyc]  & [bin cyc]   &                  &  Comment  \\
  \hline	  
260	&	50390,5	&	0,1	&	2370	&	0,72	&	20,53	&	-0,58	&	 ASM	&	25 \\
261	&	50425,4	&	0,05	&	2391	&	0,25	&	20,53	&	-0,49	&	 ASM	&	25 \\
262	&	50460,3	&	0,05	&	2411	&	0,78	&	21,00	&	-0,46	&	 ASM	&	25 \\
263	&	50496,0	&	0,1	&	2432	&	0,78	&	20,70	&	0,04	&	 ASM	&	25 \\
264	&	50531,2	&	0,4	&	2453	&	0,48	&	20,17	&	0,06	&	 ASM	&	25 \\
265	&	50565,5	&	0,2	&	2473	&	0,66	&	20,41	&	-0,09	&	 ASM	&	25 \\
266	&	50600,2	&	1,3	&	2494	&	0,06	&	20,94	&	0,47	&	 ASM	&	25 \\
267	&	50635,8	&	1,0	&	2515	&	0,00	&	20,47	&	0,91	&	 ASM	&	25 \\
268	&	50670,6	&	0,4	&	2535	&	0,47	&	20,29	&	0,47	&	 ASM	&	25 \\
269	&	50705,1	&	0,1	&	2555	&	0,76	&	20,53	&	0,02	&	  RXTE PCA &	31*\\
270	&	50740,0	&	0,2	&	2576	&	0,29	&	19,88	&	0,05	&	 ASM	&	25 \\
271	&	50773,8	&	0,3	&	2596	&	0,17	&	20,53	&	-0,57	&	 ASM	&	25 \\
272	&	50808,7	&	2	&	2616	&	0,70	&	19,94	&	-0,54	&	 ASM	&	25 \\
273	&	50842,6	&	0,3	&	2636	&	0,64	&	20,53	&	-1,10	&	 ASM	&	25 \\
274	&	50877,5	&	0,2	&	2657	&	0,17	&	20,47	&	-1,08	&	 ASM	&	25 \\
275	&	50912,3	&	0,2	&	2677	&	0,64	&	20,53	&	-1,11	&	 ASM	&	25 \\
276	&	50947,2	&	0,4	&	2698	&	0,16	&	20,47	&	-1,08	&	 ASM	&	25 \\
277	&	50982,0	&	0,2	&	2718	&	0,63	&	20,59	&	-1,11	&	 ASM	&	25 \\
278	&	51017,0	&	0,5	&	2739	&	0,22	&	19,94	&	-1,02	&	 ASM	&	25 \\
279	&	51050,9	&	0,5	&	2759	&	0,16	&	20,53	&	-1,59	&	 ASM	&	25 \\
280	&	51085,8	&	0,1	&	2779	&	0,68	&	20,53	&	-1,56	&	 ASM	&	25 \\
281	&	51120,7	&	0,5	&	2800	&	0,21	&	20,06	&	-1,53	&	 ASM	&	25 \\
282	&	51154,8	&	0,1	&	2820	&	0,27	&	20,00	&	-1,97	&	 ASM	&	25 \\
283	&	51188,8	&	0,5	&	2840	&	0,27	&	19,82	&	-2,48	&	 ASM	&	25 \\
284	&	51222,5	&	0,7	&	2860	&	0,09	&	20,40	&	-3,15	&	 ASM	&	25 \\
291	&	51465,3	&	1,0	&	3002	&	0,90	&	18,90	&	-3,84	&	 ASM, AL3	&	25 \\
296	&	51626,0	&	5,0	&	3097	&	0,42	&	18,92	&	-11,82	&	 ASM, AL3	&	25 \\
299	&	51722,5	&	1,5	&	3154	&	0,18	&	20,29	&	-16,57	&	 ASM, AL3	&	25 \\
300	&	51757,0	&	1,0	&	3174	&	0,47	&	20,35	&	-17,18	&	 ASM, AL3	&	25 \\
301	&	51791,6	&	1,0	&	3194	&	0,82	&	20,47	&	-16,92	&	 ASM, AL3	&	25 \\
302	&	51826,4	&	0,5	&	3215	&	0,29	&	20,00	&	-16,95	&	 ASM	&	25 \\
303	&	51860,4	&	0,5	&	3235	&	0,29	&	20,29	&	-17,46	&	 ASM	&	25 \\
304	&	51894,9	&	0,5	&	3255	&	0,58	&	19,82	&	-17,49	&	 ASM	&	25 \\
305	&	51928,6	&	0,3	&	3275	&	0,40	&	20,35	&	-18,34	&	 ASM	&	25 \\
306	&	51963,2	&	0,5	&	3295	&	0,75	&	19,94	&	-18,49	&	 ASM	&	25 \\
307	&	51997,1	&	0,5	&	3315	&	0,69	&	20,53	&	-19,05	&	 ASM	&	25 \\
308	&	52032,0	&	0,5	&	3336	&	0,22	&	21,06	&	-19,02	&	 ASM	&	25 \\
309	&	52067,8	&	0,5	&	3357	&	0,27	&	20,41	&	-18,47	&	 ASM	&	25 \\
310	&	52102,5	&	0,5	&	3377	&	0,68	&	21,00	&	-18,56	&	 ASM	&	25 \\
311	&	52138,2	&	0,5	&	3398	&	0,68	&	21,06	&	-18,06	&	 ASM	&	25 \\
312	&	52174,0	&	0,5	&	3419	&	0,74	&	20,59	&	-17,50	&	 ASM	&	25 \\
313	&	52209,0	&	0,5	&	3440	&	0,32	&	20,53	&	-17,42	&	 ASM	&	25 \\
314	&	52243,9	&	0,5	&	3460	&	0,85	&	20,53	&	-17,39	&	 ASM	&	25 \\
315	&	52278,8	&	0,5	&	3481	&	0,38	&	20,41	&	-17,36	&	 ASM	&	25 \\
316	&	52313,5	&	0,5	&	3501	&	0,79	&	20,88	&	-17,45	&	 ASM	&	25 \\
317	&	52349,0	&	0,5	&	3522	&	0,67	&	20,53	&	-17,07	&	 ASM	&	25 \\
318	&	52383,9	&	0,5	&	3543	&	0,20	&	20,53	&	-17,04	&	 ASM	&	25 \\
319	&	52418,8	&	0,5	&	3563	&	0,72	&	20,94	&	-17,02	&	 ASM	&	25 \\
320	&	52454,4	&	0,5	&	3584	&	0,66	&	20,47	&	-16,58	&	 ASM	&	25 \\
321	&	52489,2	&	0,5	&	3605	&	0,13	&	20,82	&	-16,61	&	 ASM	&	25 \\
322	&	52524,6	&	0,1	&	3625	&	0,95	&	20,39	&	-16,29	&	 ASM	&	25 \\
323	&	52559,3	&	0,1	&	3646	&	0,34	&	20,74	&	-16,40	&	 ASM	&	25 \\
324	&	52594,5	&	0,1	&	3667	&	0,08	&	20,91	&	-16,16	&	 ASM	&	25 \\
325	&	52630,1	&	0,1	&	3687	&	0,99	&	21,07	&	-15,75	&	 ASM	&	25 \\
326	&	52665,9	&	0,5	&	3709	&	0,06	&	20,65	&	-15,00	&	 ASM	&	25 \\
327	&	52701,0	&	0,5	&	3729	&	0,71	&	20,53	&	-15,03	&	 ASM	&	25 \\
328	&	52735,9	&	0,5	&	3750	&	0,24	&	20,94	&	-15,01	&	 ASM	&	25 \\
329	&	52771,5	&	0,5	&	3771	&	0,17	&	20,59	&	-14,57	&	 ASM	&	25 \\
330	&	52806,5	&	0,5	&	3791	&	0,76	&	20,47	&	-14,48	&	 ASM	&	25 \\
  \hline
\end{tabular} 
\vspace*{1mm} \\
\end{table*}

\begin{table*}
\vspace*{2mm}
\addtocounter{table}{-1}
\caption{continued (4)}
\label{bigtable_ext4}
\centering
\begin{tabular}{l l l l l l l l l}
\hline
35\,d       &  Turn-On    &  +/-     & orbital            & orbital      & length of           & (O - C)     &instrument  & Reference / \\
cycle no. &  [MJD]        &  [d]     & cycle no.       & phase       & cycle [bin cyc]  & [bin cyc]   &                  &  Comment  \\
  \hline	  
331	&	52841,3	&	0,5	&	3812	&	0,23	&	20,53	&	-14,51	&	 ASM	&	25 \\
332	&	52876,2	&	0,6	&	3832	&	0,76	&	20,35	&	-14,49	&	 ASM	&	25 \\
333	&	52910,8	&	0,6	&	3853	&	0,11	&	20,53	&	-14,63	&	 ASM	&	25 \\
334	&	52945,7	&	0,5	&	3873	&	0,63	&	20,86	&	-14,61	&	 ASM, AL4	&	25 \\
340	&	53158,5	&	1	&	3998	&	0,80	&	20,06	&	-12,44	&	 ASM, AL4	&	25 \\
341	&	53192,6	&	1	&	4018	&	0,86	&	20,41	&	-12,89	&	 ASM	&	25 \\
342	&	53227,3	&	0,5	&	4039	&	0,27	&	20,76	&	-12,98	&	 ASM	&	25 \\
343	&	53262,6	&	0,5	&	4060	&	0,03	&	20,23	&	-12,65	&	 ASM	&	25 \\
344	&	53297,0	&	0,5	&	4080	&	0,26	&	20,00	&	-13,04	&	 ASM	&	25 \\
346	&	53365,0	&	0,5	&	4120	&	0,26	&	20,41	&	-13,98	&	 ASM	&	25 \\
347	&	53399,7	&	0,5	&	4140	&	0,67	&	20,53	&	-14,07	&	 ASM	&	25 \\
348	&	53434,6	&	0,5	&	4161	&	0,19	&	20,47	&	-14,05	&	 ASM	&	25 \\
349	&	53469,4	&	0,2	&	4181	&	0,66	&	20,47	&	-14,08	&	 ASM, BAT	&	25, 26 \\
350	&	53504,2	&	0,3	&	4202	&	0,13	&	20,59	&	-14,11	&	 ASM, BAT	&	25, 26 \\
351	&	53539,2	&	0,3	&	4222	&	0,72	&	21,00	&	-14,02	&	 ASM, BAT	&	25, 26 \\
352	&	53574,9	&	0,1	&	4243	&	0,72	&	20,41	&	-13,53	&	 ASM, BAT	&	25, 26 \\
353	&	53609,6	&	0,5	&	4264	&	0,13	&	20,53	&	-13,62	&	 ASM, BAT	&	25, 26 \\
354	&	53644,5	&	0,1	&	4284	&	0,65	&	20,47	&	-13,59	&	 ASM, BAT	&	25, 26 \\
355	&	53679,3	&	0,1	&	4305	&	0,12	&	20,59	&	-13,62	&	 ASM, BAT	&	25, 26 \\
356	&	53714,3	&	0,1	&	4325	&	0,71	&	20,53	&	-13,53	&	 ASM, BAT	&	25, 26 \\
357	&	53749,2	&	0,1	&	4346	&	0,24	&	20,06	&	-13,51	&	 ASM, BAT	&	25, 26 \\
358	&	53783,3	&	0,7	&	4366	&	0,29	&	20,12	&	-13,95	&	 ASM, BAT	&	25, 26 \\
359	&	53817,5	&	0,1	&	4386	&	0,41	&	20,47	&	-14,33	&	 ASM, BAT	&	25, 26 \\
360	&	53852,3	&	0,1	&	4406	&	0,88	&	20,47	&	-14,37	&	 ASM, BAT	&	25, 26 \\
361	&	53887,1	&	0,2	&	4427	&	0,35	&	20,88	&	-14,40	&	 ASM, BAT	&	25, 26 \\
362	&	53922,6	&	0,5	&	4448	&	0,23	&	20,94	&	-14,02	&	 ASM, BAT	&	25, 26 \\
363	&	53958,2	&	0,2	&	4469	&	0,16	&	20,53	&	-13,58	&	 ASM, BAT	&	25, 26 \\
364	&	53993,1	&	0,1	&	4489	&	0,69	&	20,88	&	-13,55	&	 ASM, BAT	&	25, 26 \\
365	&	54028,6	&	0,1	&	4510	&	0,57	&	20,65	&	-13,17	&	 ASM, BAT	&	25, 26 \\
366	&	54063,7	&	0,2	&	4531	&	0,22	&	20,47	&	-13,14	&	 ASM, BAT	&	25, 26 \\
367	&	54098,5	&	0,1	&	4551	&	0,69	&	20,47	&	-13,06	&	 ASM, BAT	&	25, 26 \\
368	&	54133,3	&	0,2	&	4572	&	0,15	&	20,00	&	-12,91	&	 ASM, BAT	&	25, 26 \\
369	&	54167,3	&	0,2	&	4592	&	0,16	&	20,58	&	-13,58	&	 ASM, BAT	&	25, 26 \\
370	&	54202,3	&	0,4	&	4612	&	0,74	&	20,06	&	-13,50	&	 ASM, BAT	&	25, 26 \\
371	&	54236,4	&	0,3	&	4632	&	0,80	&	20,06	&	-13,95	&	 ASM, BAT	&	25, 26 \\
372	&	54270,5	&	0,2	&	4652	&	0,85	&	20,47	&	-14,39	&	 ASM, BAT	&	25, 26 \\
373	&	54305,3	&	0,2	&	4673	&	0,32	&	20,53	&	-14,42	&	 ASM, BAT	&	25, 26 \\
374	&	54340,2	&	1	&	4693	&	0,85	&	19,88	&	-14,39	&	 ASM, BAT	&	25, 26 \\
375	&	54374,0	&	0,5	&	4713	&	0,73	&	20,00	&	-15,01	&	 ASM, BAT	&	25, 26 \\
376	&	54408,0	&	0,1	&	4733	&	0,73	&	20,53	&	-15,52	&	 ASM, BAT	&	25, 26 \\
377	&	54442,9	&	0,2	&	4754	&	0,25	&	20,59	&	-15,49	&	 ASM, BAT	&	25, 26 \\
378	&	54477,9	&	0,7	&	4774	&	0,84	&	20,53	&	-15,40	&	 ASM, BAT	&	25, 26 \\
379	&	54512,8	&	0,3	&	4795	&	0,37	&	20,06	&	-15,37	&	 ASM, BAT	&	25, 26 \\
380	&	54546,9	&	0,1	&	4815	&	0,42	&	20,59	&	-15,82	&	 ASM, BAT	&	25, 26 \\
381	&	54581,9	&	0,2	&	4836	&	0,01	&	20,35	&	-15,73	&	 ASM, BAT	&	25, 26 \\
382	&	54616,5	&	0,2	&	4856	&	0,36	&	20,94	&	-15,88	&	 ASM, BAT	&	25, 26 \\
383	&	54652,1	&	0,2	&	4877	&	0,30	&	20,41	&	-15,44	&	 ASM, BAT	&	25, 26 \\
384	&	54686,8	&	0,2	&	4897	&	0,71	&	20,41	&	-15,53	&	 ASM, BAT	&	25, 26 \\
385	&	54721,5	&	0,3	&	4918	&	0,12	&	20,35	&	-15,62	&	 ASM, BAT	&	25, 26 \\
386	&	54756,1	&	0,5	&	4938	&	0,47	&	20,41	&	-15,77	&	 ASM, BAT	&	25, 26 \\
387	&	54790,8	&	0,2	&	4958	&	0,88	&	20,88	&	-15,86	&	 ASM, BAT	&	25, 26 \\
388	&	54826,3	&	0,2	&	4979	&	0,76	&	20,88	&	-15,48	&	 ASM, BAT	&	25, 26 \\
389	&	54861,8	&	0,5	&	5000	&	0,64	&	20,41	&	-15,10	&	 ASM, BAT	&	25, 26 \\
390	&	54896,5	&	0,1	&	5021	&	0,05	&	20,47	&	-15,19	&	 ASM, BAT	&	25, 26 \\
391	&	54931,3	&	0,1	&	5041	&	0,52	&	20,88	&	-15,22	&	 ASM, BAT	&	25, 26 \\
392	&	54966,8	&	0,2	&	5062	&	0,40	&	20,82	&	-14,84	&	 ASM, BAT	&	25, 26 \\
393	&	55002,2	&	0,1	&	5083	&	0,22	&	20,50	&	-14,52	&	 ASM, BAT	&	25, 26 \\
394	&	55037,1	&	0,05	&	5103	&	0,72	&	20,56	&	-14,52	&	 ASM, BAT	&	25, 26 \\
395	&	55072,0	&	0,1	&	5124	&	0,28	&	20,94	&	-14,47	&	 ASM, BAT	&	25, 26 \\
396	&	55107,6	&	0,2	&	5145	&	0,22	&	20,47	&	-14,03	&	 ASM, BAT	&	25, 26 \\
  \hline
\end{tabular} 
\vspace*{1mm} \\
\end{table*}

*
\begin{table*}
\vspace*{2mm}
\addtocounter{table}{-1}
\caption{continued (5)}
\label{bigtable_ext5}
\centering
\begin{tabular}{l l l l l l l l l}
\hline
35\,d       &  Turn-On    &  +/-     & orbital            & orbital      & length of           & (O - C)     &instrument  & Reference / \\
cycle no. &  [MJD]        &  [d]     & cycle no.       & phase       & cycle [bin cyc]  & [bin cyc]   &                  &  Comment  \\
  \hline	  
397	&	55142,4	&	0,2	&	5165	&	0,68	&	20,59	&	-14,06	&	 ASM, BAT	&	25, 26 \\
398	&	55177,4	&	0,1	&	5186	&	0,27	&	20,94	&	-13,97	&	 ASM, BAT	&	25, 26 \\
399	&	55213,0	&	0,5	&	5207	&	0,21	&	20,47	&	-13,53	&	 ASM, BAT	&	25, 26 \\
400	&	55247,8	&	0,1	&	5227	&	0,68	&	20,47	&	-13,56	&	 ASM, BAT	&	25, 26 \\
401	&	55282,6	&	0,2	&	5248	&	0,15	&	20,65	&	-13,60	&	 ASM, BAT	&	25, 26 \\
402	&	55317,7	&	0,2	&	5268	&	0,79	&	20,65	&	-13,57	&	ASM, BAT, Maxi	&	25, 26 \\
403	&	55352,8	&	0,2	&	5289	&	0,44	&	21,00	&	-13,31	&	 ASM, BAT	&	25, 26 \\
404	&	55388,5	&	0,2	&	5310	&	0,43	&	20,76	&	-12,92	&	 ASM, BAT, Maxi	&	25, 26 \\
405	&	55423,8	&	0,2	&	5331	&	0,20	&	21,00	&	-12,49	&	ASM, BAT	&	25, 26 \\
406	&	55459,5	&	0,1	&	5352	&	0,20	&	20,06	&	-12,05	&	Maxi	&	25, 26 \\
407	&	55493,6	&	0,2	&	5372	&	0,25	&	20,00	&	-12,43	&	 BAT	&	25, 26 \\
408	&	55527,6	&	0,2	&	5392	&	0,25	&	20,47	&	-12,99	&	BAT	&	25, 26, 27 \\
409	&	55562,4	&	0,2	&	5412	&	0,72	&	20,41	&	-13,02	&	 ASM, BAT, Maxi	&	25, 26, 27 \\
410	&	55597,1	&	0,2	&	5433	&	0,13	&	20,41	&	-13,11	&	 ASM, BAT, Maxi	&	25, 26, 27 \\
411	&	55631,8	&	0,7	&	5453	&	0,54	&	20,59	&	-13,20	&	 ASM, BAT, Maxi	&	25, 26, 27 \\
412	&	55666,8	&	0,1	&	5474	&	0,12	&	20,47	&	-13,12	&	 ASM, BAT, Maxi	&	25, 26, 27 \\
413	&	55701,6	&	0,1	&	5494	&	0,59	&	20,59	&	-13,15	&	 ASM, BAT, Maxi	&	25, 26, 27 \\
414	&	55736,6	&	0,1	&	5515	&	0,18	&	21,00	&	-13,06	&	BAT, Maxi	&	26, 27 \\
415	&	55772,3	&	0,5	&	5536	&	0,18	&	20,53	&	-12,56	&	BAT, Maxi	&	26, 27 \\
416	&	55807,2	&	0,1	&	5556	&	0,70	&	20,53	&	-12,54	&	BAT, Maxi	&	26, 27 \\
417	&	55842,1	&	0,1	&	5577	&	0,23	&	20,41	&	-12,51	&	BAT, Maxi	&	26, 27 \\
418	&	55876,8	&	0,4	&	5597	&	0,64	&	20,59	&	-12,60	&	BAT, Maxi	&	26, 27 \\
419	&	55911,8	&	0,1	&	5618	&	0,23	&	20,47	&	-12,51	&	BAT, Maxi	&	26, 27 \\
420	&	55946,6	&	0,3	&	5638	&	0,70	&	20,47	&	-12,55	&	BAT, Maxi	&	26, 27 \\
421	&	55981,4	&	0,05	&	5659	&	0,17	&	20,41	&	-12,58	&	BAT, Maxi	&	26, 27 \\
422	&	56016,1	&	0,2	&	5679	&	0,58	&	19,94	&	-12,67	&	BAT, Maxi	&	26, 27 \\
423	&	56050,0	&	0,2	&	5699	&	0,51	&	20,29	&	-13,23	&	BAT	&	26, 27 \\
424	&	56084,5	&	0,1	&	5719	&	0,81	&	20,88	&	-13,44	&	BAT, Maxi	&	26, 27 \\
425	&	56120,0	&	0,1	&	5740	&	0,69	&	20,12	&	-13,11	&	BAT	&	26, 27 \\
426	&	56154,2	&	0,1	&	5760	&	0,80	&	20,47	&	-13,50	&	BAT, Maxi	&	26, 27 \\
427	&	56189,0	&	0,1	&	5781	&	0,27	&	21,06	&	-13,47	&	BAT, Maxi	&	26, 27 \\
428	&	56224,8	&	0,2	&	5802	&	0,33	&	20,41	&	-12,91	&	BAT, Maxi	&	26, 27 \\
429	&	56259,5	&	0,1	&	5822	&	0,74	&	20,53	&	-13,00	&	BAT, Maxi	&	26, 27 \\
430	&	56294,4	&	0,1	&	5843	&	0,26	&	20,35	&	-12,98	&	BAT, Maxi	&	26, 27 \\
431	&	56329,0	&	0,2	&	5863	&	0,62	&	20,65	&	-13,13	&	BAT, Maxi	&	26, 27 \\
432	&	56364,1	&	0,2	&	5884	&	0,26	&	20,12	&	-13,10	&	BAT, Maxi	&	26, 27 \\
433	&	56398,3	&	0,3	&	5904	&	0,38	&	21,00	&	-13,37	&	BAT, Maxi	&	26, 27 \\
434	&	56434,0	&	0,2	&	5925	&	0,37	&	20,35	&	-12,87	&	BAT, Maxi	&	26, 27 \\
435	&	56468,6	&	0,1	&	5945	&	0,73	&	20,53	&	-12,96	&	Maxi	&	26, 27 \\
436	&	56503,5	&	0,1	&	5966	&	0,25	&	20,06	&	-12,99	&	BAT, Maxi	&	26, 27 \\
437	&	56537,6	&	0,1	&	5986	&	0,31	&	20,00	&	-13,08	&	BAT, Maxi	&	26, 27 \\
438	&	56571,6	&	0,1	&	6006	&	0,31	&	19,94	&	-13,99	&	BAT, Maxi	&	26, 27 \\
439	&	56605,5	&	0,1	&	6026	&	0,25	&	20,53	&	-14,02	&	BAT	&	26, 27 \\
440	&	56640,4	&	0,2	&	6046	&	0,77	&	20,53	&	-14,47	&	BAT, Maxi	&	26, 27 \\
441	&	56675,3	&	0,1	&	6067	&	0,30	&	20,94	&	-14,44	&	BAT, Maxi	&	26, 27 \\
442	&	56710,9	&	0,6	&	6088	&	0,24	&	20,47	&	-14,00	&	BAT, Maxi	&	26, 27 \\
443	&	56745,7	&	0,1	&	6108	&	0,71	&	20,59	&	-14,03	&	BAT, Maxi	&	26, 27 \\
444	&	56780,7	&	0,1	&	6129	&	0,30	&	21,00	&	-13,95	&	BAT, Maxi	&	26, 27 \\
445	&	56816,4	&	0,1	&	6150	&	0,29	&	20,94	&	-13,45	&	BAT, Maxi	&	26, 27 \\
446	&	56852,0	&	0,1	&	6171	&	0,23	&	20,59	&	-13,01	&	BAT, Maxi	&	26, 27 \\
447	&	56887,0	&	0,1	&	6191	&	0,82	&	19,94	&	-12,92	&	BAT, Maxi	&	26, 27 \\
448	&	56920,9	&	0,1	&	6211	&	0,76	&	20,41	&	-13,48	&	BAT, Maxi	&	26, 27 \\
449	&	56955,6	&	0,2	&	6232	&	0,17	&	20,47	&	-13,57	&	BAT, Maxi	&	26, 27 \\
450	&	56990,4	&	0,1	&	6252	&	0,64	&	20,53	&	-13,61	&	BAT, Maxi	&	26, 27 \\
451	&	57025,3	&	0,1	&	6273	&	0,16	&	20,59	&	-13,58	&	BAT, Maxi	&	26, 27 \\
452	&	57060,3	&	0,4	&	6293	&	0,75	&	20,00	&	-13,49	&	BAT, Maxi	&	26, 27 \\
453	&	57094,3	&	0,1	&	6313	&	0,75	&	20,47	&	-13,99	&	BAT, Maxi	&	26, 27 \\
454	&	57129,1	&	0,1	&	6334	&	0,22	&	20,94	&	-14,03	&	BAT, Maxi	&	26, 27 \\
455	&	57164,7	&	0,2	&	6355	&	0,16	&	20,53	&	-13,59	&	BAT, Maxi	&	26, 27 \\
456	&	57199,6	&	0,2	&	6375	&	0,68	&	20,00	&	-13,56	&	BAT, Maxi	&	26, 27 \\
  \hline
\end{tabular} 
\vspace*{1mm} \\
\end{table*}

 \begin{table*}
\vspace*{2mm}
\addtocounter{table}{-1}
\caption{continued (6)}
\label{bigtable_ext6}
\centering
\begin{tabular}{l l l l l l l l l}
\hline
35\,d       &  Turn-On    &  +/-     & orbital            & orbital      & length of           & (O - C)     &instrument  & Reference / \\
cycle no. &  [MJD]        &  [d]     & cycle no.       & phase       & cycle [bin cyc]  & [bin cyc]   &                  &  Comment  \\
  \hline	  
457	&	57233,6	&	0,1	&	6395	&	0,68	&	20,00	&	-14,06	&	BAT, Maxi	&	26, 27 \\
458	&	57267,6	&	0,5	&	6415	&	0,68	&	20,53	&	-14,56	&	BAT, Maxi	&	26, 27 \\
459	&	57302,5	&	0,1	&	6436	&	0,21	&	20,53	&	-14,54	&	BAT, Maxi	&	26, 27 \\
460	&	57337,4	&	0,1	&	6456	&	0,73	&	20,47	&	-14,51	&	BAT, Maxi, AL5	&	26, 27 \\
461	&	57372,2	&	0,1	&	6477	&	0,20	&	20,00	&	-14,54	&	BAT, Maxi	&	26, 27 \\
462	&	57406,2	&	0,1	&	6497	&	0,20	&	20,47	&	-15,04	&	BAT, Maxi	&	26, 27 \\
463	&	57441,0	&	0,1	&	6517	&	0,67	&	20,59	&	-15,07	&	BAT, Maxi	&	26, 27 \\
464	&	57476,0	&	0,1	&	6538	&	0,26	&	20,94	&	-14,99	&	BAT, Maxi	&	26, 27 \\
465	&	57511,6	&	0,1	&	6559	&	0,19	&	20,82	&	-14,55	&	BAT, Maxi	&	26, 27 \\
466	&	57547,0	&	0,3	&	6580	&	0,02	&	20,70	&	-14,23	&	BAT, Maxi	&	26, 27 \\
467	&	57582,2	&	0,1	&	6600	&	0,72	&	20,53	&	-14,02	&	BAT, Maxi	&	26, 27 \\
468	&	57617,1	&	0,1	&	6621	&	0,25	&	21,00	&	-14,00	&	BAT, Maxi	&	26, 27 \\
469	&	57652,8	&	0,1	&	6642	&	0,25	&	20,88	&	-13,50	&	BAT, Maxi	&	26, 27 \\
470	&	57688,3	&	0,1	&	6663	&	0,13	&	20,59	&	-13,12	&	BAT, Maxi	&	26, 27 \\
471	&	57723,3	&	0,3	&	6683	&	0,71	&	20,94	&	-13,03	&	BAT, Maxi	&	26, 27 \\
472	&	57758,9	&	0,3	&	6704	&	0,65	&	20,65	&	-12,59	&	BAT, Maxi	&	26, 27 \\
473	&	57794,0	&	0,2	&	6725	&	0,30	&	20,53	&	-12,56	&	BAT	&	26, 27 \\
474	&	57828,9	&	0,4	&	6746	&	0,82	&	19,94	&	-12,18	&	Maxi	&	26, 27 \\
475	&	57862,8	&	0,2	&	6765	&	0,76	&	20,06	&	-12,98	&	BAT, Maxi	&	26, 27 \\
476	&	57896,9	&	0,1	&	6785	&	0,82	&	20,06	&	-13,42	&	BAT, Maxi	&	26, 27 \\
477	&	57931,0	&	0,1	&	6805	&	0,88	&	20,41	&	-13,87	&	BAT, Maxi	&	26, 27 \\
478	&	57965,7	&	0,2	&	6826	&	0,29	&	19,94	&	-13,96	&	BAT, Maxi	&	26, 27 \\
479	&	57999,6	&	0,1	&	6846	&	0,23	&	19,94	&	-14,52	&	BAT, Maxi	&	26, 27 \\
480	&	58033,5	&	0,1	&	6866	&	0,16	&	20,06	&	-15,08	&	BAT, Maxi	&	26, 27 \\
481	&	58067,6	&	0,2	&	6886	&	0,22	&	20,17	&	-15,52	&	BAT, Maxi	&	26, 27 \\
482	&	58101,9	&	0,1	&	6906	&	0,40	&	20,35	&	-15,85	&	BAT, Maxi	&	26, 27 \\
483	&	58136,5	&	0,1	&	6926	&	0,75	&	20,53	&	-16,00	&	BAT, Maxi	&	26, 27 \\
484	&	58171,4	&	0,2	&	6947	&	0,27	&	19,88	&	-16,03	&	Maxi	&	26, 27 \\
485	&	58205,2	&	0,2	&	6967	&	0,15	&	20,06	&	-16,53	&	Maxi	&	26, 27 \\
486	&	58239,3	&	0,1	&	6987	&	0,21	&	20,70	&	-17,03	&	BAT, Maxi	&	26, 27 \\
487	&	58274,5	&	0,2	&	7007	&	0,92	&	19,88	&	-16,83	&	Maxi	&	26, 27 \\
488	&	58308,3	&	0,1	&	7027	&	0,80	&	20,12	&	-17,51	&	BAT, Maxi	&	26, 27 \\
489	&	58342,5	&	0,2	&	7047	&	0,91	&	20,59	&	-17,83	&	BAT, Maxi	&	26, 27 \\
490	&	58377,5	&	2,0	&	7068	&	0,50	&	20,06	&	-17,75	&	BAT, Maxi	&	26, 27 \\
491	&	58411,6	&	0,2	&	7088	&	0,55	&	20,59	&	-18,19	&	BAT, Maxi	&	26, 27 \\
492	&	58446,6	&	0,3	&	7109	&	0,14	&	20,53	&	-18,10	&	BAT, Maxi	&	26, 27 \\
493	&	58481,5	&	0,2	&	7129	&	0,67	&	20,53	&	-18,07	&	BAT, Maxi	&	26, 27 \\
494	&	58516,4	&	0,1	&	7150	&	0,20	&	20,53	&	-18,05	&	BAT, Maxi	&	26, 27 \\
495	&	58551,3	&	0,5	&	7170	&	0,72	&	20,00	&	-18,02	&	BAT, Maxi	&	26, 27 \\
496	&	58585,3	&	0,2	&	7190	&	0,72	&	20,47	&	-18,52	&	BAT, Maxi	&	26, 27 \\
497	&	58620,1	&	0,1	&	7211	&	0,19	&	19,94	&	-18,55	&	BAT, Maxi	&	26, 27 \\
498	&	58654,0	&	0,2	&	7231	&	0,13	&	20,00	&	-19,11	&	BAT, Maxi	&	26, 27 \\
499	&	58688,0	&	0,4	&	7251	&	0,13	&	20,53	&	-19,62	&	BAT, Maxi	&	26, 27 \\
500	&	58722,9	&	1	&	7271	&	0,65	&	20,59	&	-19,59	&	BAT, Maxi	&	26, 27 \\
501	&	58757,9	&	0,3	&	7292	&	0,24	&	20,53	&	-19,50	&	BAT, Maxi	&	26, 27 \\
502	&	58792,8	&	0,3	&	7312	&	0,77	&	20,88	&	-19,48	&	BAT, Maxi	&	26, 27 \\
503	&	58828,3	&	0,1	&	7333	&	0,65	&	20,65	&	-19,09	&	BAT, Maxi	&	26, 27 \\
504	&	58863,4	&	0,5	&	7354	&	0,29	&	20,47	&	-18,95	&	BAT, Maxi	&	26, 27 \\
505	&	58898,2	&	0,5	&	7374	&	0,76	&	20,94	&	-18,98	&	BAT, Maxi	&	26, 27 \\
506	&	58933,8	&	0,1	&	7395	&	0,70	&	20,53	&	-18,54	&	BAT, Maxi	&	26, 27 \\
507	&	58968,7	&	0,1	&	7416	&	0,23	&	20,35	&	-18,51	&	BAT, Maxi	&	26, 27 \\
508	&	59003,3	&	0,2	&	7436	&	0,58	&	20,06	&	-18,66	&	BAT, Maxi	&	26, 27 \\
509	&	59037,4	&	0,1	&	7456	&	0,64	&	20,53	&	-19,11	&	BAT, Maxi	&	26, 27 \\
510	&	59072,3	&	0,2	&	7477	&	0,16	&	20,41	&	-19,08	&	BAT, Maxi	&	26, 27 \\
511	&	59107,0	&	0,2	&	7497	&	0,57	&	20,06	&	-19,17	&	BAT, Maxi	&	26, 27 \\
512	&	59141,1	&	0,2	&	7517	&	0,63	&	20,00	&	-19,61	&	BAT, Maxi	&	26, 27 \\
513	&	59175,1	&	0,2	&	7537	&	0,63	&	20,59	&	-20,12	&	BAT, Maxi	&	26, 27 \\
514	&	59210,1	&	0,1	&	7558	&	0,21	&	20,53	&	-20,03	&	BAT, Maxi	&	26, 27 \\
515	&	59245,0	&	1	&	7578	&	0,74	&	20,47	&	-20,00	&	BAT, Maxi	&	26, 27 \\
516	&	59279,8	&	0,2	&	7599	&	0,21	&	20,12	&	-20,03	&	BAT, Maxi	&	26, 27 \\
  \hline
\end{tabular} 
\vspace*{1mm} \\
\end{table*}

 \begin{table*}
\vspace*{2mm}
\addtocounter{table}{-1}
\caption{continued (7)}
\label{bigtable_ext7}
\centering
\begin{tabular}{l l l l l l l l l}
\hline
35\,d       &  Turn-On    &  +/-     & orbital            & orbital      & length of           & (O - C)     &instrument  & Reference / \\
cycle no. &  [MJD]        &  [d]     & cycle no.       & phase       & cycle [bin cyc]  & [bin cyc]   &                  &  Comment  \\
  \hline	  
517	&	59314,0	&	0,2	&	7619	&	0,33	&	20,53	&	-20,42	&	BAT, Maxi	&	26, 27 \\
518	&	59348,9	&	0,5	&	7639	&	0,85	&	20,94	&	-20,74	&	BAT	&	26, 27 \\
519	&	59384,5	&	0,2	&	7660	&	0,79	&	20,47	&	-19,95	&	BAT, Maxi	&	26, 27 \\
520	&	59419,3	&	0,3	&	7681	&	0,26	&	20,06	&	-19,98	&	BAT, Maxi	&	26, 27 \\
521	&	59453,4	&	0,2	&	7701	&	0,32	&	20,53	&	-20,43	&	BAT, Maxi	&	26, 27 \\
522	&	59488,3	&	0,2	&	7721	&	0,84	&	20,47	&	-20,40	&	BAT, Maxi	&	26, 27 \\
523	&	59523,1	&	0,1	&	7742	&	0,31	&	20,47	&	-20,43	&	BAT, Maxi	&	26, 27 \\
524	&	59557,9	&	0,3	&	7762	&	0,78	&	20,94	&	-20,46	&	BAT, Maxi	&	26, 27 \\
525	&	59593,5	&	0,1	&	7783	&	0,72	&	20,70	&	-20,02	&	BAT	&	26, 27 \\
526	&	59628,7	&	0,2	&	7804	&	0,43	&	20,41	&	-19,94	&	Maxi	&	26, 27 \\
527	&	59663,4	&	0,2	&	7825	&	0,83	&	19,94	&	-19,56	&	BAT	&	26, 27 \\
528	&	59697,3	&	0,3	&	7844	&	0,77	&	20,41	&	-20,47	&	BAT, Maxi	&	26, 27 \\
529	&	59732,0	&	0,1	&	7865	&	0,18	&	20,00	&	-20,62	&	BAT	&	26, 27 \\
530	&	59766,0	&	0,2	&	7885	&	0,18	&	20,59	&	-21,18	&	 BAT, Maxi	&	26, 27 \\
531	&	59801,0	&	0,2	&	7905	&	0,77	&	20,06	&	-21,68	&	Maxi	&	26, 27 \\
532	&	59835,1	&	0,3	&	7926	&	-0,18	&	20,53	&	-21,18	&	Maxi	&	26, 27 \\
533	&	59870,0	&	0,2	&	7946	&	0,35	&	20,53	&	-21,57	&	Maxi	&	26, 27 \\
534	&	59904,9	&	0,5	&	7966	&	0,88	&	20,59	&	-21,36	&	BAT, Maxi	&	26, 27 \\
535	&	59939,9	&	0,2	&	7987	&	0,47	&	20,41	&	-21,28	&	BAT, Maxi	&	26, 27 \\
536	&	59974,6	&	0,1	&	8007	&	0,88	&	20,35	&	-21,37	&	BAT, Maxi	&	26, 27 \\
537	&	60009,2	&	0,1	&	8028	&	0,23	&	20,00	&	-21,52	&	BAT, Maxi	&	26, 27 \\
538	&	60043,2	&	0,1	&	8048	&	0,22	&	20,59	&	-22,02	&	BAT, Maxi	&	26, 27 \\
539	&	60078,2	&	0,1	&	8068	&	0,81	&	20,70	&	-21,93	&	BAT, Maxi	&	26, 27 \\
540	&	60113,4	&	0,2	&	8089	&	0,51	&	20,35	&	-21,73	&	BAT, Maxi	&	26, 27 \\
541	&	60148,0	&	0,2	&	8109	&	0,87	&	20,37	&	-21,88	&	BAT, Maxi	&	26, 27 \\
542	&	60182,63	&	0,03	&	8130	&	0,23	&	20,10	&	-22,09	&	BAT, ART-XC	&	26, 27, 30 \\
543	&	60216,8	&	0,1	&	8150	&	0,33	&	20,35	&	-22,41	&	Maxi	&	26, 27 \\
544	&	60251,4	&	0,2	&	8171	&	-0,32	&	20,53	&	-22,03	&	BAT	&	26, 27 \\
545	&	60286,3	&	0,1	&	8191	&	0,21	&	19,53	&	-22,53	&	Maxi	&	26, 27 \\
546	&	60319,5	&	0,6	&	8211	&	0,74	&	20,41	&	-23,09	&	BAT	&	26, 27 \\
547	&	60354,20	&	0,2	&	8230	&	0,15	&	20,06	&	-24,07	&	BAT	&	26, 27 \\
548	&	60388,3	&	0,8	&	8250	&	0,20	&	20,65	&	-24,51	&	Maxi	&	26, 27 \\
549	&	60423,4	&	0,5	&	8271	&	0,85	&	20,53	&	-23,95	&	Maxi	&	26, 27 \\
550	&	60458,3	&	0,1	&	8292	&	0,38	&	20,70	&	-23,87	&	BAT	&	26, 27 \\
551	&	60493,5	&	0,1	&	8313	&	0,08	&	20,60	&	-23,66	&	BAT, Maxi	&	26, 27 \\
552	&	60528,5	&	0,1	&	8333	&	0,68	&	20,54	&	-23,56	&	BAT, Maxi	&	26, 27 \\
553	&	60563,5	&	0,05	&	8354	&	0,22	&	20,02	&	-23,52	&	BAT, Maxi	&	26, 27 \\
554	&	60597,5	&	0,1	&	8374	&	0,25	&	20,53	&	-24,00	&	BAT, Maxi	&	26, 27 \\
555	&	60632,4	&	0,2	&	8394	&	0,77	&	19,96	&	-23,97	&	BAT, Maxi	&	26, 27 \\
556	&	60666,3	&	0,2	&	8414	&	0,74	&	20,04	&	-24,51	&	BAT	&	26, 27 \\
557	&	60700,4	&	0,1	&	8434	&	0,77	&	20,88	&	-24,97	&	BAT	&	26, 27 \\
558	&	60735,9	&	0,7	&	8455	&	0,66	&	19,65	&	-24,65	&	Maxi	&	26, 27 \\
559	&	60769,3	&	0,1	&	8475	&	0,30	&	20,88	&	-25,50	&	BAT	&	26, 27 \\
560	&	60804,8	&	0,1	&	8496	&	0,18	&	20,06	&	-24,94	&	BAT, Maxi	&	26, 27 \\
561	&	60838,9	&	0,1	&	8516	&	0,24	&	19,94	&	-25,51	&	BAT	&	26, 27 \\
562	&	60872,8	&	0,5	&	8536	&	0,18	&	20,17	&	-26,07	&	BAT, Maxi	&	26, 27 \\
563	&	60907,1	&	0,5	&	8556	&	0,35	&	20,70	&	-26,39	&	BAT, Maxi	&	26, 27 \\
564	&	60942,3	&	0,1	&	8577	&	0,06	&	20,12	&	-26,19	&	BAT, Maxi	&	26, 27 \\
565	&	60976,5	&	0,1	&	8597	&	0,17	&	20,06	&	-26,57	&	BAT, Maxi	&	26, 27 \\
566	&	61010,6	&	0,2	&	8617	&	0,23	&	20,06	&	-27,02	&	BAT, Maxi	&	26, 27 \\
567	&	61044,7	&	1	&	8637	&	0,28	&	20,06	&	-27,46	&	extrapol., AL6	&	here  \\
568	&	61078,8	&	0,2	&	8657	&	0,34	&	19,12	&	-27,90	&	NuSTAR PP$^{5}$, AL6	&	here  \\
569	&	61111,3	&	0,2	&	8676	&	0,46	&	19,00	&	-29,29	&	NuSTAR PP$^{5}$, AL6	&	here  \\
570	&	61143,6	&	0,2	&	8695	&	0,46	&		        &	-30,79	&	NuSTAR PP$^{5}$, AL6	&	here  \\
  \hline
\end{tabular} 
\vspace*{1mm} \\
$^{1}$ The numbering of 35\,d cycles follows \citet{Staubert_etal83}; 
$^{2}$  Numbers of binary cycles just before turn-on: the center of eclipse of binary cycle 0 is at MJD 46359,87194,
P(orb) = 1.70016759\,d, dP(orb)/dt = -4.85E-11\,s/s \citep{Staubert_etal09}; 
$^{3}$ "O" stands for "observed", "C" stands for "calculated"/expected; reference is  (O - C) = 0 at cycle \#31 (MJD 42409.9) 
and a constant cycle length of 20.5 x P(orb) = 34,8534356 \,d; 
$^{4}$ mean length of 35-day cycle (difference of adjacent MJD entries, divided by difference in cycle number), 
$^{5}$ from pulse profiles at different 35-day phases (see text). \\
\end{table*}

 \begin{table*}
\addtocounter{table}{-1}
\caption{continued (8)}
\label{Ref_bigtable}
{\em References}:
1 \citet{Staubert_etal83}, 2 \citet{Ulmer_etal73}, 3 \citet{Giacconi_etal73},
4 \citet{CrosaBoynton_80}, 5 \citet{McClintock_etal74}, 
6 \citet{Thomas_etal83}, 7 \citet{DavisonFabian_77}, 8 \citet{CookePage_75},
9 \citet{Pravdo_etal79}, 10 \citet{Parmar_etal80}, 11 \citet{Soong_etal90},
12 \citet{Soong_88}, 13 \citet{Gorecki_etal82}, 14 \citet{Soong_etal87}, 
15 \citet{VrtilekHalpern_85}, 16 \citet{Ohashi_etal84}, 17 \citet{Delgado_etal83}, 
18 \citet{Ushimaru_etal89}, 19 \citet{Oegelman_etal85},
20 \citet{Kahabka_87}, 21 \citet{Mihara_etal91}, 22 \citet{Scott_93},
23 Voges 1992, priv. communication, \\
24 CGRO/BATSE: 
https://astro-gdt.readthedocs.io/en/latest/missions/
cgro/batse/finders.html\#finding-continuous-batse-data, \\
25 RXTE/ASM: http://xte.mit.edu/ASM\_lc.html, \\
26 Swift/BAT: http://swift.gsfc.nasa.gov, \\
27 Maxi: http://maxi.riken.jp, \\
28 Fermi/GBM:  https://gammaray.nsstc.nasa.gov/gb,m/science/pulsars.html, \\
29 \citet{Baykal_etal93}, 30 \citet{Revnivtsev_etal25}, 31 \citet{Kuster_etal05}.  \\
*  turn-on times have been read off published graphs. \\
+ corrected when obvious mistakes in referencing time had been identified
(e.g. Ref. 7, Fig. 2: the turn-on of cycle 0 is one day earlier!).
\end{table*}

\end{document}